\font\bfs=cmb10 at 7pt

\font\bb=msbm10 at 10pt
\font\bbT=msbm10 at 8pt

\font\cal=cmsy10 at 9pt
\font\cals=cmsy10 at 8pt

\font\rms=cmr10 at 8pt

\font\sf=cmss10 at 10pt 

\def\0#1{\mbox{\rm#1}}
\def\1#1{\mbox{\bb#1}}
\def\2#1{\mbox{\bf#1}}
\def\3#1{{\cal #1}}
\def\4#1{\mbox{\cals#1}}
\def\5#1{\mbox{\sf#1}}
\def\6#1{\mbox{\rms #1}}
\def\7#1{\mbox{\bfs #1}}
\def\8#1{{\tilde #1}}
\def\9#1{{\breve #1}}

\def\BEq{\begin{equation}}
\def\EEq{\end{equation}}
\def\BEqA{\begin{eqnarray}}
\def\EEqA{\end{eqnarray}}
\def\BEn{\begin{enumerate}}
\def\EEn{\end{enumerate}}

\def\Cbb{\mbox{\bb C}}
\def\Cbbs{\mbox{\bbT C}}

\def\Fbb{\mbox{\bb F}}

\def\Mbb{\mbox{\bb M}}

\def\Rbb{\mbox{\bb R}}
\def\Rbbs{\mbox{\bbT R}}

\def\tav{\hbox{
\kern-1pt\rule[0pt]{1.5pt}{.8pt}{\kern-3.3pt}
\rule[0pt]{.4pt}{5pt}{\kern-4.4pt}
\rule[5pt]{4.5pt}{.8pt}{\kern-3.45pt}
\rule[0pt]{.4pt}{5.3pt}{\kern-1pt}
}}

\def\Ac{{\cal A}}
\def\Bc{{\cal B}}
\def\Cc{{\cal C}}

\def\Ic{{\cal I}}
\def\Lc{{\cal L}}

\def\Pc{{\cal P}}

\def\Rc{{\cal R}}

\def\Crm{{\rm C}}

\def\Irm{{\rm I}}
\def\Irms{\mbox{\scriptsize\rm I}}

\def\Orm{{\rm O}}

\def\Srm{{\rm S}}
\def\Srms{{\rms S}}
\def\Srms{\mbox{\rms S}}

\def\Urm{{\rm U}}

\def\Cb{{\bf C}}

\def\Rb{{\bf R}}

\def\D{\Delta}

\def\f{\phi}

\def\g{\gamma}

\def\ga{\gamma}

\def\m{\mu}

\def\S{\Sigma}

\def\si{\sigma}
\def\Si{\Sigma}

\def\y{\psi}

\def\BOX{\mathop{\hbox{\vrule height 4pt width 4pt depth
0pt}}}

\def\dag{\dagger}

\def\adj{{^{\dag}}}

\def\ox{\otimes}

\def\Ox{\bigotimes}

\def\pa{\partial}

\def\from{\kern-2pt\leftarrow\kern-2pt}

\def\x{\times}

\def\bra{\langle}

\def\ket{\rangle}

\def\II{|\kern-1pt |}

\def\lar{\leftarrow}

\def\uar{\uparrow}

\def\Av{\mathop{\hbox{\rm Av}}\nolimits}

\def\Cliff{\mathop{\hbox{\rm Cliff}}\nolimits}

\def\End{\mathop{\rm End}\nolimits}

\def\Seq{\mathop{\hbox{\rm Seq}}\nolimits}

\def\Set{\mathop{\hbox{\rm Set}}\nolimits}

\def\Sib{\mathop{\hbox{\rm Sib}}\nolimits}

\def\SO{\mathop{\mbox{\rm SO}}\nolimits}

\def\SU{{\mbox{\rm SU}}}

\def\Ten{\mathop{\hbox{\rm Ten}}\nolimits} 

\def\tav{\hbox{
\kern-1.0pt
\rule[0pt]{1.3pt}{.8pt}{\kern-3.6pt}
\rule[0pt]{.4pt}{6pt}{\kern-3.0pt}
\rule[4.5pt]{3.0pt}{.8pt}{\kern-3.3pt}
\rule[0pt]{.4pt}{5pt}{\kern-1pt}
}}

\def\BOX{\mathop{\hbox{\vrule height 4pt width 4pt depth
0pt}}}

\def\adj{{^{\dag}}}

\def\dag{\dagger}

\def\ox{\otimes}

\def\Ox{\bigotimes}

\def\pa{\partial}

\def\from{\kern-2pt\leftarrow\kern-2pt}

\def\x{\times}

\def\lmult{{\lfloor\kern-5pt\lfloor}}
\def\rmult{{\rfloor\kern-5pt\rfloor}}

\def\lar{\leftarrow}

\def\uar{\uparrow}

\def\adj{{^{\dag}}}
\def\dag{\dagger}

\def\ox{\otimes}
\def\Ox{\bigotimes}

\def\bra{\langle}

\def\ket{\rangle}

\def\II{|\kern-1pt |}

\def\GeV{\mbox{ GeV}}

\def\psid{{\dot \psi}}
\def\phid{{\dot \phi}}

\def\ad{{\dot a}}

\def\bd{{\dot b}}

\def\ed{{\dot e}}

\def\fd{{\dot f}}

\def\Gd{{\dot G}}

\def\Vd{{\dot V}}

\documentclass[12pt]{article}

\usepackage{makeidx,graphics,latexsym}

\begin{document}

\title{{\bf Quantum theory of elementary processes}}

\author{ \\ \\ A THESIS \\
Presented to \\
The Faculty of the Division of Graduate Studies \\
by \\
Andrei A.
Galiautdinov
\\ \\
In Partial Fulfillment \\
of the Requirements for the Degree \\
Doctor of Philosophy in Physics
\\ \\
\normalsize {\it School of Physics, Georgia Institute of
Technology, Atlanta, Georgia 30332-0430 }}

\date{\today}

\pagenumbering{roman}

\pagestyle{empty}

\maketitle

\newpage

\pagestyle{plain}

\tableofcontents

\newpage

\section*{\center  SUMMARY}

\addcontentsline{toc}{section}{SUMMARY}

In modern physics, one of the greatest divides is that
between space-time and quantum fields, as
the fiber bundle of the Standard Model indicates. However on
the operational grounds the fields and space-time are not
very different. To describe
a field in an experimental region we have to assign
coordinates to the points of that region in order to speak
of "when" and "where" of the field itself. But to
operationally study the topology and to coordinatize the
region of space-time, the use of radars (to send and
receive electromagnetic signals) is required. Thus the
description of fields (or, rather, processes) and the
description of space-time are indistinguishable at the
fundamental level. Moreover, classical general relativity
already says --- albeit preserving the fiber bundle
structure --- that space-time and matter are intimately
related. All this indicates that a new theory of elementary
processes (out of which all the usual processes of creation,
annihilation and propagation, and consequently the topology
of space-time itself would be constructed) has to be devised.

In this thesis the foundations of such a finite,
discrete, algebraic, quantum theory are presented. The
theory is then applied to the description of spin-1/2 quanta
of the Standard Model.

\newpage

\pagenumbering{arabic}

\section*{ \center INTRODUCTION}

\addcontentsline{toc}{section}{INTRODUCTION}

{\it...One can give good reasons why reality
cannot at all be represented by a continuous field. From
the quantum phenomena it appears to follow with
certainty that a finite system of finite energy can be
completely described by a finite set of numbers (quantum
numbers). This does not seem to be in accordance with a
continuum theory, and must lead to an attempt to find
a purely algebraic theory for the description of
reality. But nobody knows how to obtain the basis of
such a theory. ---ALBERT EINSTEIN
\cite{EINSTEINTHEMEANING}}

\vskip10pt

There are essentially three major areas of physics that have
been puzzling me for the last several
years: possibility of quantum-field-space-time unification
beyond the Standard Model, its experimental tests, and the
cosmological constant problem. All of them are very closely
related, and reflect our, physicists', belief that Nature at
its deepest and most fundamental level is simple, is
governed by simple universal laws, and that there must be a
theory that could describe, explain, and unify all the
intricacies of the world around us in one beautiful and
elegant scheme. I do, however, believe that
no such ``final'' theory of Nature can ever be found, at
least not the one based on mathematical axiomatic method, as
Goedel's theorems indicate (Appendix A).

The quest for unification originated in my numerous
conversations with David Finkelstein, who had always
emphasized the importance of algebraic simplicity in
physics. This simplicity is exactly what mathematicians mean
when they speak of simple groups and algebras. In physics,
algebraic simplicity is the symbol of unity and beauty. For
example, in classical mechanics, if time couples into space
(as is expressed by Galilean transformations), then there
should be --- on the grounds of simplicity --- a coupling of
space into time, and that's exactly what Lorentz
transformations establish.

In modern physics, one of the greatest divides is that
between space-time and quantum fields, as
the fiber bundle of the Standard Model indicates. However on
the operational grounds the fields and space-time are not
very different. To describe
a field in an experimental region we have to assign
coordinates to the points of that region in order to speak
of "when" and "where" of the field itself. But to
operationally study the topology and to coordinatize the
region of space-time, the use of radars (to send and
receive electromagnetic signals) is required. Thus the
description of fields (or, rather, processes) and the
description of space-time are indistinguishable at the
fundamental level. Moreover, classical general relativity
already says --- albeit preserving the fiber bundle
structure --- that space-time and matter are intimately
related. All this indicates that a new theory of elementary
processes (out of which all the usual processes of creation,
annihilation and propagation, and consequently the topology
of space-time itself would be constructed) has to be devised.

In this thesis I present the foundations of such a
finite, discrete, algebraic, quantum theory, and apply it to
the description of spin-1/2 quanta of the Standard Model.
The basic principle of the theory can be summarized as
follows:

{\bf 1) Operationality.}

The statements of physics have the form: {\bf If we do
so-and-so, we will find such-and-such}. The primary element
of the theory is thus a {\it process} (called {\it
operation} when driven by the experimenter).

{\bf 2) Process atomism.}

Any process,
dynamical or not, of any physical system must be viewed as
an {\it aggregate} of isomorphic elementary operations of
finite duration $\chi$, provisionally called {\it chronons}.

{\bf 3)  Algebraic simplicity.}

The dynamical and the symmetry groups (and, consequently, the
operator algebra) of any physical system
must be {\it simple} (in algebraic sense).

{\bf 4) Clifford statistics.}

Chronons obey Clifford statistics.

\vskip25pt

The operator algebra of a
chronon aggregate is a real Clifford algebra of a very large
number of dimensions. The chronons themselves are
represented by the generators (spins)
$\gamma_n$ of that Clifford algebra, with the
property
$\gamma_n^2=\pm 1$.

   Using the algebra of an ensemble of many chronons I
algebraically simplify the non-semisimple Dirac-Heisenberg
algebra of relativistic quantum mechanics, unifying the
space-time, energy-momentum, and spin variables of the
electron. I also propose a new dynamics that reduces to
Dirac's dynamics for fermions with the usual Heisenberg
commutation relations in the continuum limit, when the
number $N$ of elementary processes becomes infinite and
their duration
$\chi$ goes to zero. The complex imaginary unit $i$, now a
dynamical variable, the mass term $m$, and a new
spin-orbit coupling not present in the Standard Model, all
appear naturally within the new simplified theory.

This thesis is based on the work done in collaboration with
David Finkelstein. It closely follows the
material presented in
\cite{FG, GF, G0201052}.

\newpage

\section*{ \center  CHAPTER I \\ \center
ELEMENTARY OPERATIONS AND CLIFFORD
STATISTICS}

\addcontentsline{toc}{section}{CHAPTER I. ELEMENTARY OPERATIONS
AND CLIFFORD STATISTICS}

\label{sec:CLIFFORDSTATISTICSCHAPTER}

{\it ...The situation, however, is somewhat as
follows. In order to give physical significance to
the concept of time, processes of some kind are
required which enable relations to be established
between different places. ---ALBERT EINSTEIN
\cite{EINSTEINTHEMEANING}}

\vskip10pt

\subsection*{PROCESS ATOMISM}
\label{sec:ATOMISMofACTIONS}

We start our work with the idea that {\bf any process,
dynamical or not, of any physical system should be viewed as
an aggregate of isomorphic elementary operations}.
We call such fundamental operations
{\it chronons}.

The
dynamics of a physical system is usually specified either by
a Hamiltonian, or by a Lagrangian. It represents the time
evolution of the system under study between our initial and
final determination actions. The simplest and more or less
representative example of a dynamical process is
that of time evolution of an electron subjected to an
external electromagnetic field.

Let us assume that a system of classical charges,
magnets, current carrying coils and wires is distributed in
a definite way throughout the experimental region. It
produces some definite electric and magnetic fields, in which
the electron under study moves. The action of the field on
the electron defines electron's dynamics.

We can change the dynamics by re-arranging
the elements of the field-producing system --- the charges,
the wires, and so on. If the dynamics is composed of
elementary operations, as is assumed by our atomistic
hypothesis, the change in the dynamics must be accomplished
by permutation of the underlying chronons.

If we want to have different dynamics, the
permutation has to have a definite effect on the aggregate
of our fundamental operations. In other words, we need to
assign a statistics to the chronons, and that statistics
must be {\it non-abelian}.\footnote{Recall that a statistics
is {\em abelian} if it represents the permutation group $S_N$
on the
$N$ members of an ensemble by an abelian group of operators
in the
$N$-body mode space.

The usual Fermi-Dirac or Bose-Einstein statistics are
abelian. In a sense they are trivially abelian because
they represent each permutation by a number, a projective
representation of the identity operator. Entities
with scalar statistics are regarded as {\it
indistinguishable}. Thus bosons and fermions are
indistinguishable.

Non-abelian statistics describe distinguishable quanta.

One non-trivial example of non-abelian statistics was given
by  Nayak and Wilczek
\cite{wilczek, NW} in their work on quantum Hall effect.
It was based on the earlier work on nonabelions of Read and
Moore
\cite{moore90, read92}. Read and Moore use a subspace
corresponding to the degenerate ground mode of some
realistic Hamiltonian as the representation space for a
non-abelian representation of the permutation group $S_{2n}$
acting on the composite of $2n$ quasiparticles in the
fractional quantum Hall effect. This statistics, Wilczek
showed, represents the permutation group on a spinor space,
permutations being represented by non-commuting
spin operators.
The quasiparticles of Read and Moore and of Wilczek and Nayak
are distinguishable,
but their permutations leave the ground subspace invariant.
}

Thus, the first
question that has to be answered in setting up an algebraic
quantum theory of composite processes is:
What
statistics do the elementary actions have?

Ordinarily, processes (space-time events, field values,
etc.) are been assumed, though implicitly, to be
distinguishable, being
addressed by their space-time coordinates. If considered
classical, they obey Maxwell-Boltzmann statistics.

In our theory, the elementary
processes of nature obey Clifford statistics, which is
similar to the spinorial statistics of Nayak and Wilczek. We
apply this statistics first to toy models of particles in
ordinary space-time simply to familiarize ourselves with
its properties. In our construction the representation
space of the permutation group is the whole (spinor) space
of the composite. The permutation group is not assumed to be
a symmetry of the Hamiltonian or of its ground subspace. It
is used not as a symmetry group but as a dynamical group of
an aggregate.

\subsection*{QUANTIFICATION PROCEDURES AND STATISTICS}

\label{sec:QUANTIFICATIONPROCEDURES}

Apart from the atomism of actions discussed above, we suggest
that {\bf at higher energies the present complex quantum theory
with its unitary group will expand into a real quantum theory
with an orthogonal group, broken by an approximate
$i$ operator at lower energies}. To implement this
possibility and to account for double-valuedness of spin in
Nature, we develop a new real quantum double-valued
statistics. In this statistics, called Clifford, we
represent a swap (12) of two quanta by the difference
$\gamma_1-\gamma_2$ of the corresponding Clifford units. The
operator algebra of an ensemble is the
Clifford algebra over a one-body real Hilbert space. Unlike
the Maxwell-Boltzmann, Fermi-Dirac, Bose-Einstein, and para-
statistics, which are tensorial and single-valued, and
unlike anyons, which are confined to two dimensions,
Clifford statistics is multivalued and works for any
dimensionality. Interestingly enough, a similar statistics
was proposed by Nayak and Wilczek \cite{NW} for the
excitations in the theory of fractional quantum Hall effect.

To develop some feel for this new statistics, we first apply
it to toy quanta. We distinguish between the two
possibilities: a real Clifford statistics and a complex one.
A complex-Clifford example describes an ensemble with the
energy spectrum of a system of spin-1/2 particles in an
external magnetic field. This (maybe somewhat prematurely)
supports the proposal that the double-valued rotations ---
spin --- seen at current energies arise from
double-valued permutations --- swap --- to be seen at higher
energies. Another toy with real Clifford statistics
illustrates how an effective imaginary unit $i$ can arise
naturally within a real quantum theory.

\subsubsection*{QUANTIFIERS}

All the common statistics, including Fermi-Dirac (F-D),
Bose-Einstein (B-E) and Maxwell-Boltzmann (M-B), may be
regarded as various prescriptions
for constructing the algebra of an
ensemble of many individuals from the vector space of one
individual.
These prescriptions convert ``yes-or-no'' questions about an
individual into ``yes-or-no'' and ``how-many'' questions
about an ensemble of similar individuals.
Sometimes these procedures are called ``second
quantization.'' This terminology is unfortunate for obvious
reasons. We will not use it in our work. Instead, we will
speak of {\it quantification}.

We use the operational formulation
of quantum theory presented in
Appendix A. As we pointed out,
kets represent initial modes
(of preparation), bras represent final modes
(of registration), and operators represent intermediate
operations on quantum. The same applies to an ensemble of
several quanta.

Each of the {\it usual} statistics may be defined
by an associated linear mapping $Q^{\dag}$ that maps any
one-body initial mode $\psi$ into a many-body creation
operator:
\BEq\label{eq:linquantification}
Q^{\dag}: V_{\Irms} \rightarrow \Ac_{\Srms}, \; \psi
\mapsto Q^{\dag}
\psi=: \hat\psi.
\EEq
Here $V_{\Irms}$ is the initial-mode vector space of the
individual $\Irm$ and $\Ac_{\Srms}=\End V_{\Srms}$ is the
operator (or endomorphism) algebra of the quantified system
$\Srm$.
The $\dag$ in $Q\adj$  reminds us that $Q\adj$
is contragredient to the initial modes $\psi$.
We write the mapping $Q\adj$ to the left of its argument
$\psi\adj$ to respect the conventional Dirac order of
cogredient and contragredient vectors in a contraction.

Dually, the final modes
$\psi^{\dag}$ of the dual space $V^{\dag}_{\Irms}$ are
mapped to annihilators in $\Ac_{\Srms}$ by the linear
operator $Q$,
\BEq
Q: V^{\dag}_{\Irms} \rightarrow
\Ac_{\Srms}, \; \psi^{\dag} \mapsto
\psi^{\dag} Q =: \hat\psi^{\dag}.
\EEq

We
call the transformation
$Q$ the {\em quantifier} for the statistics.
$Q$
and $Q^\dag$ are tensors of the type
\BEq
Q=(Q^{aB}{}_C),\quad Q^\dag =({{Q\adj_a}^C}_B),
\EEq
where $a$ indexes a basis in the one-body space $V_{\Irms}$
and $B, C$ index a basis in the many-body
space $V_{\Srms}$.

The basic creators and annihilators associated
with an { arbitrary}
basis
$\{ e_a | \, a=1,\dots, N\} \subset V_{\Irm} $ and its
reciprocal basis $\{ e^{a} | \, a =1,\dots, N\} \subset
V^\dag_{\Irm}$ are then
\BEq
\label{eq:BASICCREATORS}
Q^{\dag} e_a := \hat e_a =: Q\adj{}_a,
\EEq
and
\BEq
\label{eq:BASICANNIHILATORS}
e^{a} Q := \hat e^{a} =: Q^{a}.
\EEq
The creator and annihilator for a general
initial mode
$\psi$ are
\BEqA
Q^{\dag} (e_a \psi^a) &= &Q\adj_a \psi^a\/,\cr
(\phi\adj_{a} e^{a}) Q &=&
\phi\adj_{a} Q^{a}\/, \EEqA
respectively.

We require that quantification respects the adjpoint $\dag$.
This relates the two tensors
$Q$ and
$Q^{\dag}$:
\BEq
\label{eq:condition}
\psi^{\dag} Q = {(Q^{\dag} \psi)}^\dag . \EEq
The rightmost $\dag$ is the adjoint operation for the
quantified system.
Therefore
\BEq
\label{eq:conditionGENERATORS}
\hat{e}_a^{\dag} = M_{a b} \hat e^{b},
\EEq
with $M_{a b}$ being the metric,
the matrix of the adjoint
operation, for the individual system.

A few remarks on the use of
creation and annihilation operators.

If we choose to work exclusively with an
$N$-body system, then all the initial and final
selective actions (projections, or yes-no experiments) on
that system can be taken as {\it simultaneous} sharp
production and registration of {\it all} the
$N$ particles in the composite with no need
for one-body creators and annihilators. The theory would
resemble that of just one particle.  The elementary
non-relativistic quantum theory of atom provides such an
example.

In real experiments much more
complicated processes occur. The number of particles
in the composite may vary, and if a special {\it vacuum}
mode is introduced, then those processes can conveniently be
described by postulating elementary operations of
one-body creation and annihilation. Using just the notion of
the vacuum mode and a simple rule by which the creation
operators act on the many-body modes, it is possible to show
(Weinberg \cite{WEINBERG1}) that {\it any} operator of
{\it such} a many-body theory may be expressed as a sum of
products of creation and annihilation operators.

In physics shifts in description are very frequent,
especially in the theory of solids. The standard example
is the phonon description of collective excitations in
crystal lattice. There the fundamental system is an ensemble
of a fixed number of ions without any special vacuum mode.
An equivalent description is in
terms of a variable number of phonons,
their creation and annihilation operators, and the
vacuum.

It is thus possible that a deeper theory underlying the
usual physics might be based on a completely new kind of
description. Finkelstein some time ago
\cite{DRFPRIVATE} suggested that the role of atomic
processes in such a theory might be played by swaps (or
permutations) of quantum space-time events. Elementary
particles then would be the excitations of a more
fundamental system. The most natural choice for the swaps is
provided by the differences of Clifford units
(\ref{eq:SWAPS}) defined in Appendix F.

All this prompts us to generalize from the common statistics
to more general statistics.

     A {\em
linear statistics} will be defined by a linear correspondence
$Q\adj$ called the quantifier,
\BEq
\label{eq:linearquantification}
Q^{\dag}: V_{\Irms} \rightarrow \Ac_{\Srms}, \; \psi
\mapsto Q^{\dag}
\psi=: \hat\psi,
\EEq
[compare (\ref{eq:linquantification})] from one-body modes
to many-body operators,
$\dag$-algebraically generating
the algebra $\Ac_{\Srms}:=\End V_{\Srms}$ of the many-body
theory.

{\bf In general $Q\adj$
does not produce a creator and $Q$ does not produce an
annihilator, as they do in the common statistics.}

\subsubsection*{CONSTRUCTING THE QUANTIFIED ALGEBRA}

We construct the quantified algebra
$\Ac_{\Srms}$ from the individual space $V_{\Irms}$
in four steps \cite{DRFPRIVATE} as follows (see
Appendix C for details.):

1. We form the quantum algebra
$\Ac(V_{\Irms})$, defined as the free $\dag$ algebra
generated by (the vectors of) $V_{\Irms}$.
Its elements are
all possible iterated sums and products and
$\dag$-adjoints of the vectors of $V_{\Irms}$.
We require that the  operations $(+, \x, \dag)$ of
$\Ac(V_{\Irms})$ agree with those of $V_{\Irms}$ where both
are meaningful.

2.  We construct the ideal $\Rc\subset \Ac$ of all elements of
$\Ac(V_{\Irms})$ that vanish in virtue of the statistics.
It is convenient and customary to define $\Rc$ by a set of
expressions $\Rb$, such that
the
commutation relations between elements of
$\Ac(V_{\Irms})$ have the form  $r=0$ with $r\in \Rb$.
Then $\Rc$ consists of all elements of
$\Ac(V_{\Irms})$ that vanish
in virtue of the commutation relations
and the postulates of a $\dag$-algebra.

Let $\Rb$ be closed under $\dag$.
Let $\Rc_0$ be the set of all
evaluations of all the expressions in $\Rb$
when the variable vectors
$\psi$ in these expressions assume any values $\psi \in
V_{\Irms}$.
Then $\Rc= \Ac(V_{\Irms}) \, \Rc_0\,
\Ac(V_{\Irms})$. Clearly, $\Rc$ is a {\it two-sided} ideal
of $\Ac(V_{\Irms})$.

3. We form the quotient algebra
\BEq
\Ac_{\Srms}=  \Ac(V_{\Irms}) /\Rc,
\EEq
by identifying elements of $\Ac(V_{\Irms})$
whose differences belong to
$\Rc$.

If one is interested in the system with a fixed number of
particles, one adds a step:

4. Take the subalgebra $P A_S P$, where $P$ is the projection
on the selected eigenspace of the {\it number operator}
$N_\Srm$,
\BEq
N_\Srm:= \sum_{a=1}^{N} \hat e_a \hat e^a,
\EEq
where $k$ labels the basis elements of $V_\Irm$ and $N = \dim
V_\Irm$.

Thus in all the usual statistics, $\Ac_\Srm$ is
the sum
\BEq
\label{eq:SUMMANYBODYALGEBRA}
\Ac_\Srm = \Ac_0 + \Ac_1 + \Ac_2 + \dots
\EEq
of 0-, 1-, 2-, ... particle algebras. In Clifford
statistics (see below), on the other hand,
the ``number operator'' is
\BEq
N_C = \sum_{a, b=1}^N g_{ab} \gamma^a \gamma^b = N,
\EEq
($\gamma^a$ being the Clifford generators), which means that
the number
$N$ of elements in the Clifford ensemble is
fixed by the dimensionality of the corresponding
one-quantum Hilbert space.

In any case, as the result of the above construction, $Q\adj$
maps each vector
$\y\in V_{\Irms}$ into its residue class
$\y+\Rc$.

Historically, physicists carried out
one special quantification first.
Since in classical physics one multiplies
phase spaces when quantifying,
they assumed that in quantum mechanics one should multiply
Hilbert spaces,
forming the
tensor product
\BEq
V_{\Srms}=\Ox_{p=0}^N V_{\Irms} =  V_{\Irms}^N
\EEq
of $N$ individual spaces
$V_{\Irms}$.
To improve agreement with experiment,
they removed
degrees of freedom in the tensor product connected with
permutations,
reducing
$V_{\Irms}^N$ to a subspace $P_\Srm V_{\Irms}^N$ invariant
under all permutations of individuals.
Here $P_\Srm$ is a
projection operator characterizing the statistics.
The many-body algebra was then taken to be the algebra of
linear operators on the reduced space: $\Ac_{\Srms}=\End
P_\Srm V_{\Irms}^N$.

We call a statistics built in that way
on a subspace of the tensor algebra over the one-body
initial mode space, a {\em
tensorial} statistics.
Tensorial statistics represents permutations in a
single-valued way. The common statistics are tensorial.

Linear statistics is more general than tensorial
statistics, in that the quotient algebra
$\Ac_{\Srms}=\Ac(V_\Irm )/\Rc$ defining a linear statistics
need not be the operator algebra of any subspace of the
tensor space $\Ten V_{\Irms}$ and need not be single-valued.
Commutation relations permit more general statistics
than projection
operators do.
For example, anyon statistics is linear but not tensorial.

For another example, $\Ac_{\Srms}$ may be the endomorphism algebra
of a spinor space constructed from the quadratic space
$V_{\Irms}$.
Such a statistics we call a {\em spinorial
statistics}\/.
Clifford statistics, the main topic of
this work,
is a spinorial statistics.
Linear statistics includes both
spinorial and tensorial statistics \cite{DRFPRIVATE}.

The  F-D, B-E and M-B statistics are readily
presented as
tensorial statistics.
We give their quantifiers next
\cite{DRF96}.
We then generalize to spinorial, non-tensorial, statistics.

\subsubsection*{STANDARD STATISTICS}
\label{sec:STANDARDSTATISTICS}

{\bf Maxwell-Boltzmann statistics.}
A classical M-B aggregate is a sequence
(up to isomorphism)
and
$Q=\Seq$, the {\it sequence}-forming quantifier.
The quantum
individual $\Irm$ has a Hilbert space
$V=V_{\Irms}$ over the field ${\Cbb}$.
The vector space for
the quantum sequence then is the (contravariant) tensor
algebra $V_{\Srms}=\Ten V_{\Irms}$, whose product is the
tensor product $\ox$:
\BEq
V_{\Srms}=\Ten V_{\Irms}
\EEq
with the natural induced $\dag$.
The kinematic algebra $\Ac_{\Srms}$ of the sequence is the
$\dag$-algebra of endomorphisms of $\Ten V_{\Irms}$, and is
generated by $\psi \in V_{\Irms}$ subject to the generating
relations
\BEq
\hat \psi\adj \hat{\phi} = \y\adj \f.
\EEq
The left-hand
side is an operator product, and the right-hand side
is the contraction of the dual vector
$\y^\dag$ with the vector $\f$, with an implicit
unit element $1\in \Ac_ {\Srm}$ as a factor.

{\bf Fermi-Dirac statistics.} Here $Q=\Set$,
the {\it
set}-forming quantifier.
The kinematic algebra for the
quantum set has defining relations
\BEqA
\hat{\y} \hat{\f} + \hat{\f}\hat \y &=&
0, \;\cr
\hat\y^{\dag}\hat \f +\hat \f \hat\y^{\dag} &=& \y\adj \f ,
\EEqA
for all $\y,\f\in V_{\Irms}$\/.

{\bf Bose-Einstein statistics.} Here $Q=\Sib$, the {\it
sib}-forming quantifier. The sib-generating relations are
\BEqA
\hat\y \hat\f -\hat \f \hat\y &=& 0, \;\cr \hat\y^{\dag}
\hat\f - \hat\f
\hat\y^{\dag} &=& \y\adj \f ,
\EEqA
for all $\y,\f\in V_{\Irms}$\/.

The individuals in each of the discussed quantifications,
by construction, have the same (isomorphic) initial spaces.
We call such individuals {\it isomorphic}.

\subsubsection*{THE REPRESENTATION PRINCIPLE}
\label{sec:REPRESENTATIONPRINCIPLE}

If we have defined how, for example, one translates
individuals, this should define a way to translate the
ensemble. We thus impose the following

{\bf Requirement of a
quantification:} {\it any unitary transformation on an
individual quantum entity induces a unitary transformation on
the quantified system, defined by the quantifier.}

This does not imply that, for example, the actual
time-translation of an ensemble
is carried out by translating the individuals:
That would
mean that the Hamiltonians combine additively,
without
interaction.
The representation principle states only that there is a
well-defined time-translation without interaction.
This gives a physical meaning to interaction: it is the
difference between the induced time translation generator
and the actual one.

Thus we posit that any $\dag$-unitary
transformation
\[
U: V_{\Irms} \rightarrow V_{\Irms}, \, \psi
\mapsto U\psi
\]
  of the individual ket-space
$V_{\Irms}$, also act naturally on the quantified mode
space $V_{\Srms}$ through an operator
\[
\hat{U}: V_{\Srms}\to V_{\Srms}, \, \psi
\mapsto \hat{U} \psi,
\]
and on the algebra $\Ac_{\Srms}$ according to
\BEqA
\hat{U}:
\Ac_{\Srms} \rightarrow \Ac_{\Srms}, \, \hat\y \mapsto
\widehat{ U\y}=\hat U \hat\y \hat{U}^{-1}.
\EEqA
This is the {\em representation principle}.

{\bf Recall:} Infinitesimal form of $U:V_{\Irms}\to
V_{\Irms}$ is
\BEq
U = 1+G\delta \theta\/,
\EEq
where $G=-{G}^{\dag}:V_{\Irms}\to V_{\Irms}$ is
the anti-Hermitian generator and $\delta
\theta$ is an infinitesimal parameter.
The infinitesimal anti-Hermitian generators $G$ make up the
Lie algebra
$d\Urm_{\Irms}$ of the unitary group $\Urm_{\Irms}$ of the
one-body theory.

In terms of generators the representation principle is
expressed by
\BEq
\label{eq:ltfgeqa}
\hat G: \hat{\psi} \mapsto \widehat{G\y} =[\hat G,
\hat{\psi}].
\EEq

\subsubsection*{THE QUANTIFIED GENERATORS}

Since
\BEq
\label{eq:ANOTHERDEFOFOPERATOR}
G =
\sum_{a, b} e_a {{G}^a}{}_{b} e^{b}
\EEq
holds by the completeness of the dual bases $e_a$ and $e^a$\/,
we express the quantified generator $\hat G$
by\footnote{According to Weinberg \cite{WEINBERG1}, this rule
for quantifying {\it observables} was first given by Heisenberg
and Pauli \cite{HEISENBERGandPAULI} in their early work on
quantum field theory.}
\BEq
\label{eq:quantification}
\hat G := Q^{\adj} G Q =
\sum_{a , b} Q\adj_a{{G}^a}{}_{b} Q^{b}
\equiv
\sum_{a, b}\hat e_a {{G}^a}{}_{b} \hat e^{b}.
\EEq

\subsubsection*{CHECKING FOR THE USUAL STATISTICS}

The representation principle holds for
the usual F-D and B-E statistics. It will also hold
for the Clifford statistics, as we show below.

{\bf Proposition:} If $Q$ is a quantifier
for F-D or B-E statistics then
\BEq
[\hat G, \, Q\adj{\psi}] = G Q\adj{\psi}
\EEq
hold for all anti-Hermitian generators $G$.

{\bf Proof:} We have
\BEqA
\label{eq:consistency}
[\hat G, \, Q\adj{\psi}] &=
&{{G}^a}_b \, \left( \hat e_a \, \hat e^b \, Q\adj{\psi} -
Q\adj{\psi} \, \hat e_a \, \hat e^b \right) \cr
&=& {{G}^a}_b \, \left( \hat e_a \, ( e^b \psi +
(-1)^{\kappa}
\, Q\adj{\psi} \, \hat e^b) - Q\adj{\psi} \, \hat e_a \,
\hat e^b \right)
\cr &=& {{G}^a}_b \, \hat e_a \, e^b \psi \cr &=& {{G}^a}_b
\, \hat e_a \,
\psi^b \cr &=& G \, Q\adj{\psi}.
\EEqA
Here $\kappa=1$ for Fermi statistics and 0 for Bose.

\subsubsection*{QUANTIFICATION AND COMMUTATOR ALGEBRAS}

If $\Ac$ is any algebra, by the commutator algebra
$\D \Ac$ of $\Ac$ we mean the Lie algebra on the elements
of $\Ac$ whose product is the commutator
$[a,b]=ab-ba$ in $\Ac$.
By the commutator algebra of a quantum system $\Irm$ we
mean that of its operator algebra $\Ac_{\Irms}$.

In the usual cases of Bose and Fermi statistics (and not in
the cases of complex and real Clifford statistics discussed
below!) the quantification rule (\ref{eq:quantification})
defines a Lie isomorphism, $\D
\Ac_{\Irms}\to \D \Ac_{\Srms}$, from the commutator algebra
of the individual
to that of the quantified system.

{\bf Proposition:}
For two (arbitrary) operators $H$ and
$P$ acting on $V_{\Irms}$,
\BEq
\label{eq:Lie}
\widehat{[H, \, P]} = [\hat H, \, \hat P].
\EEq

{\bf Proof:}
\BEqA
[\hat H, \, \hat P] &= & \hat H \hat P - \hat P \hat H \cr
&= & \hat e_r{{H}^r}_s\hat e^s \;
\hat e_t{{P}^t}_u\hat e^u
-\hat e_t{{P}^t}_u\hat e^u \;
\hat e_r{{H}^r}_s\hat e^s \cr
&= & {{H}^r}_s {{P}^t}_u
(\hat e_r \hat e^s \hat e_t \hat e^u
-\hat e_t \hat e^u \hat e_r \hat e^s) \cr &= & {{H}^r}_s
{{P}^t}_u (\hat e_r (\delta_t^s \pm \hat e_t \hat e^s) \hat
e^u -\hat e_t \hat e^u
\hat e_r \hat e^s) \cr
&= & {{H}^r}_s {{P}^t}_u
(\hat e_r \delta_t^s \hat e^u \pm \hat e_r \hat e_t \hat
e^s \hat e^u -\hat e_t \hat e^u \hat e_r \hat e^s) \cr
&= & {{H}^r}_s {{P}^t}_u
(\hat e_r \delta_t^s \hat e^u \pm \hat e_t \hat e_r \hat
e^u \hat e^s -\hat e_t \hat e^u \hat e_r \hat e^s) \cr
&= & {{H}^r}_s {{P}^t}_u
(\hat e_r \delta_t^s \hat e^u \pm \hat e_t (\mp \delta^u_r
\pm \hat e^u
\hat e_r) \hat e^s -\hat e_t \hat e^u \hat e_r \hat e^s) \cr
&= & {{H}^r}_s {{P}^t}_u
(\hat e_r \delta_t^s \hat e^u - \hat e_t \delta^u_r \hat
e^s) \cr &= & \hat e_r ( {{H}^r}_t {{P}^t}_u -
{{P}^r}_t {{H}^t}_u) \hat e^u \cr
&= & \widehat{[H, \, P]}.
\EEqA

This implies that for B-E and F-D statistics,
the
quantification rule (\ref{eq:quantification})
can be extended from the
unitary operators and their anti-Hermitian generators to
the whole operator algebra (including observables) of the
quantified system.

\subsection*{CLIFFORD STATISTICS}

\subsubsection*{CLIFFORD QUANTIFICATION}
\label{sec:CS}

Now let the one-body mode space $V_{\Irms} =
{\Rbb}^{N_+,N_-}=N_+\Rbb
\oplus N_-\Rbb$
be a real quadratic space of dimension $N=N_+ + N_-$ and
signature
$N_+-N_-$.
Denote the symmetric metric form of $V_{\Irms}$
by $g=(g_{ab}):=(e_a^\dag e_b)$.
We do not assume that $g$ is positive-definite.

{\bf Clifford statistics}
(\ref{eq:linearquantification}) {\bf is defined by:}

(1) the Clifford-like generating relations
\BEqA
\label{eq:spinorialquantification}
\hat\y \hat\f+\hat\f \hat\y &=&  \, \frac{\zeta}{2} \,
\y\adj \f
\EEqA
for all $\f, \y\in V_{\Irms}$,
where $\zeta$ is a $\pm$ sign that can have either value;

(2) the Hermiticity condition (\ref{eq:condition})
\BEq
\label{eq:conditionGENERATORSCLIFFORD}
\hat{e}_a^{\dag} = g_{a b} \hat e^{b};
\EEq

(3) a rule for raising and lowering indices
\BEq
\label{eq:conditionGENERATORSCLIFFORD}
\hat{e}_a:= \zeta' \, g_{a b} \hat e^{b},
\EEq
where $\zeta'$ is another $\pm$ sign,
and

(4) the rule (\ref{eq:quantification}) to quantify
one-body generators.

Here $\zeta = \pm 1$ covers
the two different conventions used in the
literature.
Later we will see that
$\zeta=\zeta'$, and that
$\zeta= \zeta' = +1$ and $\zeta=\zeta' = -1$ are both
allowed physically at the present theoretical stage
of development.
They lead to two
different real quantifications, with either Hermitian or
anti-Hermitian Clifford units.

For the quantified
basis elements of $V_{\Irm}$
(\ref{eq:spinorialquantification}) leads to
\BEqA
\hat e_a \hat e_b + \hat e_b \hat e_a &=&
\frac{\zeta}{2} \, g_{ab}.
\EEqA

The $\y$'s, which are assigned grade 1
and taken to be either Hermitian or anti-Hermitian, generate
a graded
$\dag$-algebra that we call the
{\it free Clifford $\dag$-algebra}
associated with
${\Rbb}^{N_+,N_-}$ and write as
$\Cliff (N_+, N_-)$ $\equiv\Cliff(N_\pm)$.

In assuming a real vector space of quantum modes
instead of a complex one,
we give up $i$-invariance but retain quantum superposition
$a\psi + b \phi$ with real coefficients.
Our theory is non-linear from the complex point of
view.\footnote{Others considered non-linear quantum theories,
but gave up real superposition as well as $i$-invariance
\cite{weinberg}.}

\subsubsection*{CLIFFORDONS AND THEIR PERMUTATIONS}
\label{sec:PERMUTATIONS}

Clifford statistics assembles its quanta, {\it cliffordons},
individually described by
vectors into a composite described by spinors,
which we call a {\em squadron}.

A cliffordon is a hypothetical quantum-physical entity,
like an electron,
not to be confused with a mathematical object like a spinor
or an operator.
We cannot describe a cliffordon completely,
but we represent our actions on a squadron of cliffordons
adequately by operators in a Clifford algebra of operators.
One
encounters cliffordons only in permuting them,
never in
creating or annihilating them as individuals.

Clifford statistics, unlike the more familiar particle
statistics \cite{berezin, feynman, negele},
provides no
creators or annihilators.
With each individual mode $e_a$ of
the quantified system they associate a Clifford unit
$\gamma_a = 2Q\adj_a$\/.

In the standard statistics there is a natural
way to represent permutations of individuals in the
$N$-body composite.
Each $N$-body ket is constructed
by successive action of $N$ creation operators on the
special vacuum mode.
Any permutation of individuals can be achieved
by permuting these creation operators in the product.
The identity and alternative representations of the
permutation group
$S_N$ in the B-E and F-D cases then follow from the defining
relations of the corresponding statistics.

In the case of Clifford statistics, some things are
different. There is still an operator associated with each
cliffordon; now it is a Clifford unit.
Permutations of cliffordons are still
represented by operators on a many-body $\dag$ space.
But
the mode space on which these operators act is now a spinor
space,
and its  basis vectors are not constructed by
creation operators acting on a special ``vacuum'' ket.
).

We may represent any swap
(transposition of two cliffordons, say 1 and 2)
by the difference of the corresponding Clifford
units
\BEq
t_{12}:=\frac{1}{\sqrt{2}}(\gamma_1 - \gamma_2).
\EEq
and represent an arbitrary permutation, which is a product of
elementary swaps,
by the product of their representations.
That is, as direct computation shows,
this defines a projective homomorphism
from $\0S_N$ into the Clifford algebra
generated by the $\gamma_k$. For details, see
Appendix F.

By definition, the number $N$
of cliffordons in a squadron is the dimensionality of the
individual initial mode space $V_{\Irms}$.
$N$ is conserved
rather trivially, commuting with every Clifford element.
We
can change this number only
by varying the dimensionality of the one-body space.
In one use of the theory,
we can do this, for example,
by changing the space-time 4-volume of the corresponding
experimental region.
Because our theory does not use creation and annihilation
operators, an initial action on the squadron represented by a
spinor $\xi$ should be viewed as some kind of spontaneous
transition condensation into a coherent mode, analogous to
the transition from the superconducting to the many-vertex
mode in a type-II superconductor. The initial mode of a set
or sib of (F-D or B-E) quanta can be regarded as a result of
possibly entangled creation operations. That of a squadron of
cliffordons cannot.

\subsubsection*{REPRESENTATION PRINCIPLE}

As with (\ref{eq:consistency}),
let us verify that definition (\ref{eq:quantification}) is
consistent in the Clifford case:
\BEqA
\label{eq:consistencyCW}
[\hat G, \, Q\adj \psi]&=
&{{G}^a}_b \, \left( \hat e_a \, \hat e^b \, Q\adj \psi -
Q\adj \psi \,
\hat e_a \, \hat e^b \right) \cr &=& \frac{1}{2} {{G}^a}_b
\, \left( \hat e_a \psi^b + \psi_a \hat e^b \right) \cr &=&
G \, Q\adj \psi.
\EEqA
This shows that $Q\adj \psi$ transforms correctly under the
infinitesimal unitary transformation of ${\Rbb}^{N_+,N_-}$
({\it cf.}
\cite{dirac74}).

\subsubsection*{CLIFFORD STATISTICS AND COMMUTATOR ALGEBRAS}
\label{sec:QUANTIFYINGOBSERVABLES}

In the usual statistics, the quantifier $Q$ is
extended from
anti-Hermitian operators to all operators.
This is not the case for Clifford quantification.
There the quantification of any symmetric operator is a
scalar, in virtue of Clifford's law. A straightforward
calculation shows that
\BEqA
[\hat H, \, \hat P] & = &
\hat H \hat P - \hat P \hat H \nonumber \\
&=& \zeta \, \zeta' \, \left( \frac{1}{2} \widehat{[H,
\, P]} +
\frac{1}{4} \left( \widehat{[P, \, H\adj]} +
\widehat{[P\adj, \, H]} \right) \right).
\EEqA

The three simplest cases are:
\begin{enumerate}

\item $H=H\adj$, $H'=H'\adj$ $\Longrightarrow$ $[\hat H,
\hat H'] = 0$;

\item $H=H\adj$, $G_1=-G_1\adj$ $\Longrightarrow$ $[\hat H,
\, \hat G_1] = 0$;

\item $G=-G\adj$, $G'=-G'\adj$ $\Longrightarrow$ $[\hat G,
\, \hat G'] = \zeta \, \zeta' \,
\widehat{[G, \, G']}$.

\end{enumerate}

      Thus Clifford quantification respects the commutation relations
for anti-Hermitian generators
if and only if $\zeta = \zeta' = +1$ or $\zeta= \zeta' =
-1$; but not for Hermitian
observables, contrary to the Bose and Fermi quantifications,
which respect both.

\subsubsection*{COMPLEX CLIFFORD STATISTICS}
\label{sec:cwstatistics}

The {\em complex} graded algebra generated by the $\y$'s
with the relations (\ref{eq:spinorialquantification})
is called the {\em complex Clifford algebra}
$\Cliff_{\Cbbs}(N)$ over
${\Rbb}^{N_+,N_-}$. It is isomorphic
to the full complex matrix algebra
${\Cbb (2^n)\otimes \Cbb (2^n)}$ for even $N=2n$, and to
the direct sum
${\Cbb(2^n)\otimes \Cbb(2^n)} \oplus {\Cbb(2^n)\otimes
\Cbb(2^n)}$ for odd
$N=2n+1$.
We regard $\Cliff_{\Cbbs}(N)$ as the kinematic algebra of
the complex Clifford composite.
As a vector space, it has dimension $2^N$.\footnote{Schur
\cite{schur} used complex spinors and complex Clifford
algebra to represent permutations some years before Cartan
used them to represent rotations. There is a fairly
widespread view that spinors may be more fundamental than
vectors, since vectors may be expressed as bilinear
combinations of spinors.
One of us took this direction in much of his
work.
Clifford statistics support the opposite view.
There a vector describes an individual, a spinor an aggregate.
Wilczek and Zee \cite{Wilczek82} seem to have been the
first to recognize
that spinors represent composites in a physical
context,
although this is implicit
in the Chevalley construction of spinors
within a Grassmann algebra \cite{DRFPRIVATE}.}

For dimension $N=3$
spinors of $\Cliff_{\Cbbs}(3)$ have as many parameters as
vectors, but for higher $N$ the number of components of the
spinors associated with $\Cliff(N_\pm)$
grows exponentially with $N$.
The physical relevance of this
irreducible double-valued (or projective) representation of
the permutation group
$S_{N}$
was recognized by Nayak and Wilczek
\cite{NW,wilczek}
in a theory of the fractional quantum Hall effect.
We call the statistics
based on
$\Cliff_{\Cbbs}(N)$ {\it complex Clifford}
statistics.

\subsubsection*{BREAKING THE $i$ INVARIANCE}

Thus we cannot construct useful Hermitian variables of a
squadron by applying the quantifier to the Hermitian
variables of the individual cliffordon.

This is closely related to fact that the real initial mode
space
$\Rbb^{N_\pm}$ of a cliffordon has no special operator
to replace the imaginary unit $i$ of the standard
complex quantum theory. The fundamental task of the
imaginary element $i$ in the algebra of complex quantum
physics is precisely to relate conserved Hermitian
observables $H$ and anti-Hermitian generators $G$
by
\BEq
\label{eq:GHI}
H=-i\hbar G.
\EEq
To perform this function exactly, the operator $i$ must
commute exactly with all observables.

The central operators $x$ and $p$ of classical mechanics
are contractions of noncentral operators $\breve{x}$
and $\breve{p}=-i\hbar \pa/\pa \breve{x}$ \cite{fs}.
In the limit of
large numbers of individuals organized coherently into
suitable condensate modes, the expanded operators of the
quantum theory contract into the central operators of the
classical theory.
Condensations produce nearly commutative variables.

Likewise we expect the central
operator $i$ to be a contraction of a non-central operator
$\breve{i}$ similarly resulting from a condensation
in a limit of large numbers.
In the simpler expanded theory,
$\breve{i}$, the correspondent of $i$, is not
central.\footnote{One clue to the nature of $\breve{i}$ and
the locus of its condensation is how the operator $i$ behaves
when we combine separate systems. Since infinitesimal
generators $G, G',
\dots $ combine by addition, the imaginaries $i, i', \dots
$ of different individuals must combine by identification
\BEq
i=i'=\dots
\EEq
for (\ref{eq:GHI}) to hold exactly,
and nearly so for (\ref{eq:GHI}) to hold nearly. The only
other variables in present physics that combine by
identification in this way are the time
$t$ of classical mechanics and the space-time coordinates
$x^\m$ of field theories. All systems in an ensemble must
have about the same $i$, just as all particles have about
the same $t$ in the usual instant-based formulation, and all
fields have about the same space-time variables $x^{\m}$ in
field theory. We identify the variables $t$ and $x^\m$ for
different systems because they are set by the experimenter,
not the system.  This suggests that the
experimenter, or more generally the environment of the
system, mainly defines the operator
$i$.
The central operators $x, p$ characterize a small system
that results from the condensation of many particles.
The central operator $i$ must result from a condensation in
the environment; we take this to be the same condensation
that forms the vacuum and the spatiotemporal structure
represented by the variables
$x^\m$ of the
standard model \cite{DRFPRIVATE}.}

The existence of this contracted $i$
ensures that at least approximately,
every Lie commutation relation
between dimensionless anti-Hermitian generators $A, B, C$
of the  standard complex quantum theory,
\BEqA
\label{eq:relation}
[A, \, B]=C,
\EEqA
corresponds to a commutation relation between Hermitian
variables $-i\hbar A$, $-i\hbar B$, $-i\hbar C$:
\BEqA
[-i\hbar A, \, -i\hbar B] = -i\hbar(-i\hbar C) . \EEqA
It also tells us that this correspondence is not exact in
nature.

St\"{u}ckelberg \cite{stuckelberg} reformulated
complex quantum mechanics in the real Hilbert space
${\Rbb}^{2N}$ of twice as many dimensions by assuming a
special real antisymmetric operator $J: \Rbb^{2N}\to
\Rbb^{2N}$ commuting with all the variables of the
system.

A real $\dag$ or Hilbert space
has no such operator.
For example, in ${\Rbb}^{2}$ the operator
\BEq
E:=
\left[\matrix{\varepsilon_1 & 0 \cr
0& \varepsilon_2}\right]
\EEq
is a symmetric operator
with an obvious spectral decomposition representing,
according to the usual interpretation, two selection
operations performed on the system, and cannot be written in
the form $G=-J \hbar E$ relating it to some antisymmetric
generator $G$ for any real antisymmetric $J$ commuting with
$E$.

           On the other hand, if we restrict ourselves to observable
operators of the form
\BEq
E':=
\left[\matrix{\varepsilon & 0 \cr
0& \varepsilon}\right],
\EEq
we can use the operator $J$,
\BEq
\label{eq:J}
J:=
\left[\matrix{0 & 1 \cr
-1& 0}\right],
\EEq
to restore the usual connection between symmetry
transformations and corresponding observables.
This restriction can be generalized to any even
number of dimensions
\cite{stuckelberg}.

\subsubsection*{BREAKDOWN OF THE EXPECTATION VALUE FORMULA}

For a system described in terms
of a general real Hilbert space there is no simple relation
of the form $G=\frac{i}{\hbar}H$ between the symmetry
generators and the observables: the usual
notions of Hamiltonian and momentum are meaningless in that
case. This amplifies
our earlier observation that Clifford quantification $A\to
\hat A$ respects the Lie commutation relations among
anti-Hermitian generators, not Hermitian observables.

           Operationally, this means that selective acts of individual
and quantified cliffordons use essentially different sets of
filters. This is not the case for complex quantum mechanics
and the usual statistics. There some important filters for
the composite are simply assemblies of filters for the
individuals.

Again, in the complex case
the expectation value formula for an assembly
\BEq
\Av X = \psi \adj X \psi/\psi\adj \psi
\EEq
is a consequence of the eigenvalue principle for
individuals, rather than an independent assumption
\cite{DRF63, DRF96}. The argument presented
in \cite{DRF63, DRF96} assumes that the
individuals over which the average is taken combine with
Maxwell-Boltzmann statistics. For highly excited systems
this is a good approximation even if the individuals have
F-D or B-E or other tensorial statistics.
It is not necessarily a good approximation for cliffordons,
which have spinorial, not tensorial, statistics.

\subsubsection*{SPIN-1/2 COMPLEX CLIFFORD MODEL}
\label{sec:morerealistic}

In this section we present a simplest possible model of a
complex Clifford composite. The resulting many-body energy
spectrum is isomorphic to that of a sequence of spin-1/2
particles in an external magnetic field.

Recall that in the usual complex quantum theory the
Hamiltonian is related to the infinitesimal time-translation
generator $G=-G\adj$ by $G=iH$. Quantifying $H$ gives the
many-body Hamiltonian. In the framework of spinorial
statistics, as discussed above, this does not work, and
quantification in principle applies to the anti-Hermitian
time-translation generator $G$, not to the Hermitian
operator $H$. Our task now is to choose a particular
generator and to study its quantified properties.

We assume an  even-dimensional real initial-mode space
$V_{\Irms}=\Rbb^{2n}$ for the quantum individual, and
consider the
dynamics with the
simplest non-trivial
time-translation generator
\begin{eqnarray}
\label{eq:G}
G:= \varepsilon
\left[\matrix{{\bf 0}_n & {\bf 1}_n \cr
-{\bf 1}_n& \hphantom{-}{\bf 0}_n}\right] \end{eqnarray}
where $\varepsilon$ is a constant energy coefficient.

The quantified time-translation generator $\hat G$ then has
the form
\begin{eqnarray}
\label{eq:hat G}
\hat{G}&:= &\sum_{l, j}^N
\hat e_l G^l{}_j \hat e^j
\cr
&= & -\varepsilon \sum_{k=1}^{n}
(\hat e_{k+n}\hat e^k -
\hat e_k\hat e^{k+n}) \cr
&= & +\varepsilon \sum_{k=1}^{n}
(\hat e_{k+n}\hat e_k -
\hat e_k\hat e_{k+n}) \cr
&= & 2 \varepsilon \sum_{k=1}^{n}\hat e_{k+n}\hat e_k \cr
&\equiv &
\frac{1}{2} \varepsilon \sum_{k=1}^{n} \gamma_{k+n}
\gamma_k. \end{eqnarray}

By Stone's theorem, the generator $\hat G$ of time
translation in the spinor space of the complex Clifford
composite of $N=2n$ individuals can be factored
into a Hermitian $H^{(N)}$ and an imaginary unit $i$ that
commutes strongly with $H^{(N)}$:
\BEq
\hat G = i H^{(N)}.
\EEq
We suppose that $H^{(N)}$ corresponds to the Hamiltonian
and seek its spectrum.

We note that by (\ref{eq:hat G}), $\hat G$ is a sum of $n$
commuting anti-Hermitian algebraically independent operators
$\gamma_{k+n} \gamma_k, k=1,2,...,n$,
$(\gamma_{k+n} \gamma_k)^{ \dagger }=-\gamma_{k+n}
\gamma_k$,
$(\gamma_{k+n} \gamma_k)^2=-1^{(N)}$.

We use the well-known $2^n \times 2^n$
complex matrix representation of the
$\gamma$-matrices  of the complex universal Clifford
algebra associated with the real quadratic space
$\Rbb^{2n}$  (see Appendix F). We can simultaneously
diagonalize the $2^n\times2^n$ matrices representing the
commuting operators $\gamma_{k+n} \gamma_k$, and use their
eigenvalues, $\pm i$, to find the spectrum $\lambda$ of
$\hat G$, and consequently of $H^{(N)}$.

        A simple calculation shows that the spectrum of $\hat
G$ consists of the eigenvalues
\BEq
\lambda_k = \frac{1}{2} \varepsilon (n - 2k)i,\quad k=0, \,
1, \, 2,\dots,
\, n\/,
\EEq
with multiplicity
\BEq
\m_k= C^n_k:=\frac{n!}{k!(n-k)!} \/.
\EEq

The spectrum the Hamiltonian $H^{(N)}$
is
\BEq
\label{eq:spectrum H}
E_k =  \frac{1}{2}(n - 2k) \varepsilon ,
\EEq
with degeneracy $\m_k$.
Thus $E_k$ ranges over the interval
\BEq
-\frac{1}{4} N \varepsilon < E < \frac{1}{4} N \varepsilon,
\EEq
in steps of $\varepsilon$, with the given degeneracies.

Thus the spectrum of the structureless
$N$-body complex Clifford composite
is the same as that of a system of $N$ spin-1/2
Maxwell-Boltzmann particles of magnetic moment $\m$ in a
magnetic field $\vec{H}$, with the identification \BEq
\frac{1}{4} \varepsilon = \mu H\/.
\EEq

Even though we started with such a simple one-body
time-translation generator as (\ref{eq:G}),  the
spectrum of the resulting many-body Hamiltonian possesses
some complexity, reflecting the fact
that the units in the composite are distinguishable, and
their swaps generate the
dynamical variables of the system.

This spin-1/2 model
does not tell us how to swap two
Clifford units experimentally.
Like the phonon model of the harmonic oscillator,
the statistics of the
individual quanta enters the picture only through the
commutation relations among the fundamental operators of the
theory.

\subsubsection*{REAL CLIFFORD STATISTICS}
\label{sec:cliffordstatistics}

According to the Periodic Table of the Spinors
\cite{budinich, lounesto, porteous, snygg},
the free (or universal) Clifford
algebra
$\Cliff_{\Rbbs}(N_+, N_-)$ is algebra-isomorphic to the
endomorphism algebra of a module $\S(N_+, N_-)$ over a
ring ${\cal R}(N_+, N_-)$.

{\bf The Periodic Table of the Spinors (here $\zeta = - 1$):}
\BEq
\label{eq:PERIODIC}
\begin{array}{l|cccccccccccccc}
&N_- &  0  &  1   &  2  &  3  &  4  &  5  &  6  &  7
&\dots&
\\ \hline
N_+&\\

0 && \1R & \1R_2 & 2\1R & 2\1C & 2\1H & 2\1H_2 & 4\1H &
8\1C &\dots
\\
1
&  &\1C & 2\1R & 2\1R_2 & 4\1R& 4\1C & 4\1H & 4\1H_2 &
8\1H&\dots
\\
2
& & \1H & \1C_2 & 4\1R & 4\1R_2 & 8\1R& 8\1C & 8\1H &
8\1H_2 &\dots
\\
3
& &\1H_2& 2\1H & 4\1C & 8\1R & 8\1R_2 & 16\1R& 16\1C &
16\1H &\dots
\\
4 & &
2\1H &2\1H_2& 4\1H & 8\1C & 16\1R & 16\1R_2 & 32\1R&
32\1C &\dots
\\
5
& &
4\1C &
4\1H &4\1H_2& 8\1H & 16\1C & 32\1R & 32\1R_2 & 64\1R &\dots
\\
6
            & & 8\1R& 8\1C &
8\1H &8\1H_2& 16\1H & 32\1C & 64\1R & 64\1R_2 &\dots
\\
7
            &  &8\1R_2
            & 16\1R& 16\1C &
16\1H &16\1H_{2}& 32\1H & 64\1C & 128\1R &\dots
\\
\vdots & &\vdots &\vdots &\vdots &\vdots &\vdots &\vdots
&\vdots &\vdots
\\
\end{array}
\EEq

The ring of coefficients
${\cal R}(N_+, N_-)$ varies periodically with period 8 in
each of the dimensionalities
$N_{+}$ and
$N_-$ of
$V_{\Irms}$, and is a function of signature $N_{+}-N_-$
alone.

In our application the module $\S(N_+, N_-)$, the spinor
space supporting
$\Cliff_{\Rbbs}(N_+, N_-)$,  serves as the initial mode
space of a squadron of $N$ real cliffordons.
${\cal R}(N_\pm)$ we call the {\em spinor
coefficient ring}
for $\Cliff_{\Rbbs}(N_+, N_-)$.

\subsubsection*{EMERGENCE OF A QUANTUM $i$ IN THE REAL CLIFFORD
STATISTICS}

The Periodic Table of the Spinors suggests another origin
for the complex $i$ of quantum theory, and one that is not approximately
central but exactly central.
Some Clifford algebras
$\Cliff_{\Rbbs}(N_+, N_-)$ have the spinor coefficient ring
$\Cbb$, containing an element $i$.
Multiplication by this $i$
then represents an  operator in the center of the
Clifford algebra, which we designate also by $i$.
We may use $i$-multiplication to represent the top element
$\g^{\uar}$ whenever $\g^{\uar}$  is central and has square $-1$.
This $i\in \Cliff_{\Rbbs}(N_\pm)$ corresponds
to the $i$ of complex quantum theory.

{\bf Examples:}

$\Cliff_{\Rbbs}(1, 0)$ is
commutative;

$\Cliff_{\Rbbs}(0,3)$ and $\Cliff_{\Rbbs}(5,0)$ are
non-commutative.

  Triads or pentads of such cliffordons could
underlie the physical ``elementary'' particles, giving rise to
complex quantum mechanics within the real.

Let us consider $\Cliff_{\Rbbs}(0, 3)=\Cbb (2)$. Its Pauli
representation is
$\gamma_1 := i \, \sigma_1, \; \gamma_2 := i
\, \sigma_2 , \; \gamma_3 := i \, \sigma_3$
with $\zeta=-1$\/.
Choose a particular one-cliffordon dynamics of
the form
\BEq
G := \left[\matrix{ 0 & V & 0  \cr -
V & 0 & \varepsilon \cr
          0& -\varepsilon &0}\right].
\EEq
Quantification (\ref{eq:quantification}) of $G$ gives
\BEq
\hat G = i \; H^{(-3)}
\EEq
with the Hamiltonian
\BEq
H^{(-3)} = \frac{1}{2}
\left[\matrix{ V & \varepsilon \cr
\varepsilon &-V}\right].
\EEq
This is also the Hamiltonian for a generic two-level
quantum-mechanical system (with the energy separation
$\varepsilon$) in an external potential field $V$, like
the ammonia molecule in a static electric field
discussed in \cite{FEYNMANLECTURES}.

\newpage

\section*{\center CHAPTER II \\ \center
ALGEBRAIC SIMPLICITY AND DIRAC'S
DYNAMICS}

\addcontentsline{toc}{section}{CHAPTER II. ALGEBRAIC SIMPLICITY
AND DIRAC'S DYNAMICS}

{\it...it is contrary to the mode of thinking
in science to conceive of a thing (the space-time
continuum) which acts itself, but which cannot be
acted upon. ---ALBERT EINSTEIN
\cite{EINSTEINTHEMEANING}}

\vskip10pt

\subsection*{ALGEBRAIC SIMPLICITY}

\subsubsection*{SIMPLE THEORIES AND UNIFICATION PROGRAMS}

A {\em simple} theory is one with simple (irreducible)
dynamical and symmetry groups. What is not simple or
semi-simple we call {\em compound}\/. A {\em contraction} of
a theory is a deformation of the theory in which some
physical scale parameter, called the {\em simplifier},
approaches a singular limit,  taken to be 0
with no loss of generality. The contraction of a simple
theory is in general compound \cite{Segal51, Inonu52, INONU}.
By {\em simplification} we mean the more creative, non-unique
inverse process, finding a simple theory that contracts to a
given compound theory and agrees better with experiment.
The main revolutions in physics of the twentieth century were
simplifications with simplifiers $c, G, \hbar$.

One sign of a compound theory is a
breakdown of reciprocity,
the principle
that every coupling works both ways.
The classic example is
Galilean relativity.
There reciprocity  between space and time
breaks down;
boosts couple time into space and there is no
reciprocal coupling.
Special relativity established reciprocity
by  replacing the compound Galilean bundle of space
fibers over the time base by the simple Minkowski space-time.
Had Galileo insisted on simplicity and reciprocity he could have formulated
special relativity in the 17th century
(unless he were to choose $\SO(4)$ instead of
$\SO(1,3)$).
Every bundle theory violates reciprocity
as much as Galileo's. The bundle group couples
the base to the fiber but not conversely.
Every bundle theory cries out for simplification.

This now requires us to establish
reciprocity
between space-time (base coordinates) $x^{\mu}$
and energy-momenta (fiber coordinates)
$p_{\mu}$.\footnote{Segal
\cite{Segal51} postulated $x\leftrightarrow p$ symmetry exactly
on grounds of algebraic simplicity;
his work stimulated that of In\"on\"u and Wigner, and ours.
Born \cite{BORN} postulated $x\leftrightarrow p$ reciprocity,
on the grounds
that it is impossible in principle
to measure the usual four-dimensional interval
of two events within an atom.
We see no law
against  measuring space-time coordinates and
intervals at that gross scale.
We use his term ``reciprocity'' in a broader sense
that includes his.}

Einstein's gravity theory and the Standard Model
of the other forces are bundle theories,
with field space as fiber and space-time as base.
Therefore these theories are
ripe for simplification \cite{fs}.
Here we simplify a spinor theory,
guided by criteria of
experimental adequacy, operationality, causality,  and
finity.

Classical field theory is but a singular limit of quantum
field theory; it suffices to simplify the quantum field
theory. Quantum field theory in turn we regard as
many-quantum theory. Its field variables all arise from spin
variables of single quanta. For example, a spinor field
arises from the theory of a single quantum of spin $1/2$ by
a transition to the many-body theory, or {\it
quantification} (the transition from the one-body to the many-body
theory,  converting yes-or-no predicates about an individual
into how-many predicates about an aggregate of isomorphic
individuals;
as distinct from quantization).
To unify field with space-time in quantum field theory,
it suffices to unify spin with space-time
in the one-quantum theory, and to quantify
the resulting theory. We unify in this thesis
and quantify in a sequel.

Some unification programs concern themselves with simplifying
just  the
internal symmetry group of the elementary particles,
ignoring the fracture between
the internal and external variables.
They attempt to unify (say) the hypercharge, isospin
and color variables,  separate from the space-time
variables.
Here we close the greatest wound first,
expecting that the
internal variables will unite with each other when
they unite with the external variables;
as uniting space with time
incidentally unified the electric and magnetic fields.
We represent space-time variables
$x^{\mu}$
and $p_{\mu}$ as
approximate descriptions of many spin variables,
in one quantum-spin-space-time structure
described in a higher-dimensional spin algebra.
This relativizes the split between field and space-time,
as Einstein relativized the split between space and
time.\footnote{A different approach to the
quantum-field-space-time unification is provided by
supersymmetry. It will not be considered in this
work.}

The resulting quantum atomistic space-time
consists of many small
exactly Lorentz-invariant isomorphic
quantum elements which we call {\em
chronons}.\footnote{Feynman, Penrose and Weizs\"acker
attempted to  atomize space or space-time into
quantum spins.
R. P.
Feynman wrote a space-time vector as the sum of a great
many Dirac spin-operator vectors \cite{FEYNMAN},
$
x^{\mu}\sim \sum_n\gamma^{\mu}(n),
$
Penrose
dissected the sphere $S^2$ into a spin network
\cite{PENROSE}; his work
inspired this program.
Weizs\"acker \cite{WEIZSAEKER},
attempted a cosmology of spin-1/2 urs.
The respective groups
are Feynman's $\SO(3,1)$, Penrose's $\SO(3)$,
Weizs\"ackers $\SU(2)$ and our
$\SO(3N, 3N)$ ($N\gg 1$).}

Simplifying a physical theory generally detaches us
from a supporting condensate.\footnote{For Galileo and
Kepler, the condensate was the Earth's crust, and to detach
from it they moved in thought to
a ship or the moon, respectively
\cite{GALILEO, KEPLERSOMNIUM}.}
In the present situation of physics the prime condensate is
the ambient vacuum.
Atomizing space-time enables us to present the vacuum as a
condensate of a simple system,
and to detach from it in thought by a phase transition,
a space-time melt-down.

\subsubsection*{CHRONON STATISTICS}

Chronons carry a fundamental time-unit
$\chi$, one of our simplifiers.
    Finkelstein (see Appendix D)
   has argued that $\chi$ is
much greater than the Planck time and is on the order of the
Higgs time $\hbar/M_H c^2$.\footnote{In
an earlier effort to dissect
space-time,  assuming multiple Fermi-Dirac
statistics for the elements \cite{DRF69, DRF96}.
This false start  led  us eventually to the Clifford-Wilczek
statistics
\cite{Wilczek82, NW, wilczek, FINKELSTEINRODRIGUEZ84, FG};
an
example of Clifford-Wilczek statistics is unwittingly
developed in Chapter 16 of \cite{DRF96}.}
We now replace the
classical Maxwell-Boltzmann statistics  of space-time events
with the simple Clifford-Wilczek statistics appropriate for
distinguishable isomorphic units.
This  enormously reduces
the problem of forming a theory.

We single out two main quantifications
in field theories like
gravitation and the Standard Model:

A classical quantification assembles a
space-time
from individual space-time points.

A separate quantification constructs a  many-quantum theory
or quantum field theory from a
one-quantum theory
on that space-time.

In the standard physics the space-time quantification tacitly
assumes Maxwell-Boltzmann  statistics for the elements of
space-time, and the field quantification uses Fermi-Dirac or
Bose-Einstein statistics.
The simplified theory we propose uses
one Clifford quantification for all of
these purposes.

In this thesis we work only with one-quantum
processes of $N\gg 1$ chronons.
To describe several quanta and their
interactions,
getting closer to field theory and experiment,
will require no further quantification,
but only additional internal combinatory structure
that is readily accomodated within the
one Clifford-Wilczek quantification.

For reader's convenience we briefly recapitulate the main
points of the previous chapter.

\paragraph{Statistics}
One defines the statistics of
an (actual, not virtual) aggregate by defining how the
aggregate transforms under permutations of its units.
That is,
to describe $N$ units with given unit
mode space  $V_1$ we give, first,
the mode space $V_N$ of the aggregate quantum system
and, second, a simple representation $R_N: \0S_N\to \End
V_N$ of the permutation group $\0S_N$ on the given $N$ units
by linear operators on
$V_N$.
This also defines the quantification
that converts yes-or-no questions
about the individual
into how-many questions about
a crowd.

In {\em Clifford statistics} $\End
V_N$ is a Clifford algebra $C=\Cliff(V_1)$,
and so $V_N$ is a spinor space for that Clifford algebra,
with $C=\End V_N$.
We write $C_1$ for the first-degree
subspace of $C$.
A Clifford statistics is defined by a
projective (double-valued) representation
$R_C: S_N\to C_1\subset C$ of the permutations
by first-grade Clifford elements
over the unit mode space $V_1$ \cite{FG}.
To define $R_C$ we associate with the $n$th unit
(for all $n=1, \dots, N$) a Clifford unit $\gamma_n$,
and we represent every swap (transposition
or 2-cycle) $(mn)$ of two distinct units by
the difference $\pm(\gamma_n-\gamma_m)\in C_1$
of the associated Clifford units.

Some useful terms:

     A {\em cliffordon} is a
quantum with Clifford statistics.

A {\em squadron} is a quantum aggregate of cliffordons.

A {\em sib} is a quantum aggregate of bosons.

A {\em set} of quanta is an
aggregate of fermions.

A {\em sequence} of quanta is a
aggregate of Maxwell-Boltzmann quanta with a given
sequential order \cite{DRF69}.

$R_C$ can be extended to a spinor
representation of $\SO(N)$
on a spinor space $\Si(N)$.

The symmetry group $G_U$ of the quantum kinematics for a universe $U$ of
$N_U$ chronons is an orthogonal group
\BEqA
G_U&=&\SO(N_{U+}, N_{U-}),\cr
N_U&=&N_{U+} +N_{U-}.
\EEqA
The algebra of observables of $U$
is the simple
finite-dimensional real Clifford algebra
\BEq
C_U=\Cliff (V_1) =
\Cliff[1, \gamma(1),\dots, \gamma(N_U)]
\EEq
generated by the  $N_U$ Clifford units
$\gamma(n), \; n= 1, \dots, N_U$ representing
exchanges.
The Clifford units $\gamma(n)$
span a vector space $V_1 \cong C_1$ of first-grade elements
of $C_U$.

Within
$C_U$ we shall construct a simplified Dirac-Heisenberg
algebra
\BEq
\breve{\Ac}_{\rm DH} =
{\Ac}[\breve{i},
\breve p, \breve x,\tilde \gamma]
\subset
C(V_1)
\EEq
whose commutator Lie algebra is simple
and which
contracts to the usual
Dirac-Heisenberg algebra
$\Ac_{\rm DH}$
in the continuum limit.
We factor  ${\breve \Ac_{\rm DH}}$
into the Clifford product
\BEq
\label{eq:CLIFFORDPRODUCT}
\breve \Ac_{\rm DH}=\Cliff(N{\bf
6_0}) =  \Cliff((N-1){\bf 6_0})\sqcup\Cliff({\bf 6_0})
\EEq
of two Clifford algebras,
an ``internal'' algebra from the last hexad
and an ``external" algebra from
all the others.

We designate our proposed simplifications of $\gamma$ and
$i, \hat p$, $\hat x$, and $\hat
O$  by  $ \tilde \gamma $ and $\breve{i},
\breve p, \breve x$ and $ \breve O$.
In the limit $\chi\to 0$
the
tildes
$\tilde{\phantom{m}}$ disappear
and the breves $\breve{\phantom{m}}$ become hats
$\hat{\phantom{m}}$.


\subsubsection*{RELATIVISTIC DIRAC-HEISENBERG ALGEBRA}

Each physical theory defines at least three algebras
that should be simple:

1) the associative operator algebra of the system \cite{DRF96,
DRF99},

2) the
kinematical Lie algebra consisting of possible Hamiltonians,
and

3) the symmetry Lie algebra of one preferred Hamiltonian.

There is no second quantization.
But there is a second simplification; and a third, and so on,
all of different kinds with different simplifiers.
Each
of the historic revolutions that guide us now
introduced a simplifier, small on the scale of
previous experience
and therefore long overlooked, into the multiplication table and
basis
elements of one or more of these algebras,
and so deformed a
compound algebra into a simpler algebra
that works better.
Among these simplifiers are $c, G$, and $\hbar$.

Here we simplify the free Dirac equation and its underlying
{\em Dirac-Heisenberg} (real unital
associative) algebra
\BEq
\label{eq:DHALGEBRA}
\Ac_{\rm DH} =\Ac_{\rm D} \ox \Ac_{\rm H}.
\EEq

Since we take the notion of dynamical process as primary in
physics \cite{DRF73}, we express a measurement
(observation, filtering, selection, yes-no experiment) as a
special case of an interaction between observer and system.
     Therefore we first simplify the
     anti-Hermitian space-time and energy-momentum symmetry
     generators
     $\hat p^\mu$ and $\hat x_\nu$,
     not the associated Hermitian observables $p^\mu$,
     $x_\nu$.
Then we simplify the Hermitian operators
by multiplying the anti-Hermitian ones by a
suitably simplified $i$.

The Dirac-Heisenberg algebra (\ref{eq:DHALGEBRA}) is a tensor
product of the Dirac and the relativistic Heisenberg
algebras, in turn defined as follows:

\paragraph{Relativistic Heisenberg algebra}
$\Ac_{\rm H}=\Ac [i,\hat p, \hat x]$ is generated
by the imaginary unit $i$ and the space-time and energy-
momentum
translation generators $\hat p_\nu := ip_\nu \equiv - \hbar
{\partial }/{ \partial x^\nu} $ and
$\hat x^\mu :=ix^\mu$, subject to
the relations
\BEqA
\label{eq:HA}
[\hat p^\mu, \hat x^\nu] &=& -i \hbar g^{\mu\nu} , \cr
[\hat p^\mu, \hat p^\nu] &=& 0, \cr
[\hat x^\mu, \hat x^\nu] &=& 0, \cr
[ i, \hat p^\mu] &=& 0, \cr
[ i, \hat x^\mu] &=& 0, \cr
i^2&=&-1.
\EEqA
Here $g^{\mu\nu}$ is the Minkowski metric, held fixed in
this paper. The hats (on $\hat{p}$, for example) indicate
that a factor $i$ has been absorbed to make the operator
anti-Hermitian \cite{adler}.
The algebra $\Ac_{\rm H}$ has both the usual associative
product and the Lie commutator product.
As a Lie algebra
$\Ac_{\rm H}$ is compound, Segal emphasized,
containing the  non-trivial
ideal generated by the unit $i$.

The
orbital Lorentz-group  generators are
\BEq
\hat O^{\mu\nu} :=
iO^{\mu\nu} = -i \left(\hat x^\mu \hat p^\nu - \hat x^\nu
\hat p^\mu
\right).
\EEq
These automatically obey the usual relations
\BEqA
[\hat O^{\mu \nu}, \hat O^{\lambda \kappa}] &=& \hbar \left(
g^{\mu \lambda} \hat O^{\nu \kappa}
- g^{\nu\lambda} \hat O^{\mu \kappa}
- g^{\mu\kappa} \hat O^{\nu\lambda}
+ g^{\nu\kappa} \hat O^{\mu\lambda} \right), \cr
[\hat x^\mu, \hat O^{\nu \lambda}] &=& \hbar \left(
g^{\mu\nu}
\hat x^\lambda - g^{\mu\lambda}
\hat x^\nu \right),\cr
[\hat p^\mu, \hat O^{\nu \lambda}] &=& \hbar
\left( g^{\mu\nu}
\hat p^\lambda -  g^{\mu \lambda}
\hat p^\nu \right),\cr
[ i, \hat  O^{\mu \nu}] &=& 0 .
\EEqA

\paragraph{Dirac algebra}
$\Ac_{\rm D}=\Ac[\gamma_{\mu}]$ is generated
by Dirac-Clifford units $\gamma_{\mu}$  subject to the
familiar relations
\BEq
\label{eq:DA}
\{\gamma_\nu , \gamma_\mu \} = 2 g_{\nu \mu}.
\EEq
As usual we write $\gamma_{\mu\nu\dots}$
for the anti-symmetric part
of the tensor $\gamma_{\mu}\gamma_{\nu}\dots$ .

\subsection*{SIMPLIFICATION OF THE RELATIVISTIC HEISENBERG
ALGEBRA}

As already mentioned, field theory employs a
compound field-space-time bundle with space-time for base
and field-space for fiber; just as Galilean space-time
is a four-dimensional bundle
with $\Rbb^3$ for base and  $\Rbb^1$ for fiber.
The prototype is the covector field,
where the fiber is the cotangent space to space-time,
with coordinates that we designate by $p_{\mu}$.

{\bf Main assumption:} {\it in experiments of sufficiently high
resolution the
space-time tangent bundle (or the
Dirac-Heisenberg algebra) manifests itself as a simple
quantum-field-space-time synthesis.}

The
space-time variables $x^\mu$ and  the tangent space
variables $p_\mu$ unite into one simple construct,
as space and time have already united.
Now, however, the simplification requires an atomization,
because the field variable actually derives from
an atomic spin.

We first split the space-time tangent bundle
into quantum cells. The minimum number of elements in a cell
for our simplification is six: four for space-time and two
for a complex or symplectic plane. We provisionally adopt the
hexadic cell.\footnote{Earlier work, done by
Finkelstein and co-workers before our present stringent
simplicity requirement, assumed a pentadic cell
\cite{FINKELSTEINRODRIGUEZ84}. This provided no natural
correspondent for the energy-momentum operators
\cite{DRFPRIVATE}.}

$N$ hexads define a unit mode space $V_1 = N {\bf 6_0}$.
There are two possibilities here as to how to proceed. We
can either use the Hexad Lemma described in Appendix E and
work with hexadic cells whose variables commute with one
another, or we can work directly with $6N$ anticommuting
Clifford units of
$\Cliff(3N,3N)$. In any case the external variables of the
quantum probe related to different hexads as defined below will
commute. The first possibility, however, might be the key to the
derivation of the Maxwell-Boltzmann statistics for classical
space-time points from a deeper quantum theory of elementary
processes.

Operationally we do not deal with empty space-time.
We explore space-time with one relativistic quantum
spin-$1\over2$ probe of rest mass  $m \sim 1/\chi$.
We express the usual spin operators $\ga^{\mu}$, space-time position operators
$x^{\mu}$, and energy-momentum  operators $p_{\mu}$
of this probe
as contractions of operators
in the Clifford algebra $
\Cliff(3N,3N)$.

We write the dynamics of the usual contracted, compound
Dirac
theory in manifestly covariant form,
with a Poincar\'e-scalar Dirac operator
\BEq
\label{eq:DIRACDYNAMICS}
D=\ga^{\mu} p_{\mu} - mc .
\EEq
$D$ belongs to the algebra of operators
on spinor-valued functions $\psi(x^\mu)$ on space-time.
Any physical spinor
$\psi(x^{\mu})$  is to obey the dynamical equation
\BEq
\label{eq:DYNAMICS}
D\psi=0.
\EEq
We simplify the dynamical operator $D$,
preserving the form of the dynamical equation
(\ref{eq:DYNAMICS}).

\subsubsection*{SIMPLIFYING EXTERNAL VARIABLES}

\label{sec:STVARIABLES}

The compound symmetry group for the Dirac equation
is the covering group of the Poincar\'e group
$\Irm{\Srm}\Orm(\Mbb)$.
We represent this as the contraction of a
simple group $\SO(3,3)$ acting on the
spinor pseudo-Hilbert (ket) space of
$6N$ Clifford generators
$\ga^{\omega}(n)$ ($\omega =0, \dots, 5$; $n=1,\dots, N$)
of the orthogonal group $\SO(3N, 3N)$.
The size of the
experiment fixes the parameter $N$.

As in Dirac one-electron theory (where
the spin generators are represented by second-degree elements
\BEq
\label{eq:DIRACGENERATORS}
\hat S^{\mu \nu}:=
\frac{\hbar}{4}\left[\gamma^\mu, \gamma^\nu\right]
\equiv
\frac{\hbar}{2}\gamma^{\mu\nu}, \quad
\mu, \nu = 0,
\dots, 3
\EEq
of the Clifford algebra $\Cliff(1,3)$),
we use the second degree elements of our Clifford algebra
$\Cliff V_1 = \Cliff(N{\bf 6_0})$ to represent anti-Hermitian
generators of rotations, boosts, space-time and energy-momentum
shifts.

We associate the position and momentum axes
with the $\ga^4$ and $\ga^5$ elements of the hexad respectively,
so that
an infinitesimal orthogonal transformation
in the 45-plane couples momentum into position.
This accounts for the symplectic symmetry of
classical mechanics and the $i$ of quantum mechanics.

{\bf Our choice of the simplified $\breve
i$, $\breve x^\mu$, and
$\breve p^\nu$ of the probe is:}
\BEqA
\breve i \equiv \frac{1}{N-1}\sum_{n=1}^{N-1}
\breve i(n) &:=& \frac{1}{N-1} \sum_{n=1}^{N-1}
\gamma^{45}(n) ,\cr\cr
\breve x^\mu \equiv \sum_{n=1}^{N-1} \breve
x^{\mu}(n) &:=& -\chi
\sum_{n=1}^{N-1}
\gamma^{\mu 4}(n) , \cr\cr
\breve p^\nu
\equiv \sum_{n=1}^{N-1} \breve
p^{\nu}(n) &:=& \phi
\sum_{n=1}^{N-1}
\gamma^{\nu 5}(n),
\EEqA
where $\chi$, $\phi$ and $N$ are simplifiers of our theory,
and
\BEq
\gamma^{\rho \sigma}(n)
:= \frac{1}{2} [\gamma^{\rho}(n), \gamma^{\sigma}(n)].
\EEq

To support this choice for the expanded generators we form
the following commutation relations among them ({\it
cf.} \cite{SNYDER, DOPLICHER}):
\BEqA
\label{eq:EXTERNALCOMMRELSINGLE}
[\breve p^\mu, \breve x^\nu] &=& -2 \phi
\chi (N-1)\;  g^{\mu\nu} \, \breve i,
\cr \cr
[\breve p^\mu, \breve p^\nu] &=& -\frac{4
\phi^2}{\hbar}
\;   \breve L^{\mu \nu} , \cr
\cr [\breve x^\mu, \breve x^\nu] &=& -\frac{4
\chi^2}{\hbar}\;
       \breve L^{\mu \nu},
\cr \cr
[\breve i, \breve p^\mu ] &=&
- \frac{2\phi}{\chi (N-1)}
\;   \breve x^\nu, \cr \cr
[\breve i, \breve x^\mu ] &=&
+ \frac{2\chi}{\phi (N-1)}
\;   \breve p^\mu.
\EEqA

     In (\ref{eq:EXTERNALCOMMRELSINGLE}),
\BEqA
\label{eq:ANGULARMOMENTA}
\breve L^{\mu\nu}&:= &\frac{\hbar}{2}
\sum_{n=1}^{N-1} \gamma^{\mu\nu}(n) ,\cr
\breve J^{\mu\nu}&:= &\frac{\hbar}{2}
\sum_{n=1}^{N} \gamma^{\mu\nu}(n)
\equiv L^{\mu\nu}+S^{\mu\nu} .
\EEqA
where $\breve S^{\mu\nu}$ is
the Dirac spin operator ({\it cf.}
(\ref{eq:DIRACSPINOPERATOR})),

$J^{\mu\nu}$
obeys the Lorentz-group commutation relations:
\BEq
[\breve J^{\mu \nu}, \breve J^{\lambda \kappa}] = \hbar
\left( g^{\mu \lambda} \breve J^{\nu \kappa}
- g^{\nu\lambda} \breve J^{\mu \kappa}
- g^{\mu\kappa} \breve J^{\nu\lambda}
+ g^{\nu\kappa} \breve J^{\mu\lambda} \right),
\EEq
and generates a total Lorentz transformation of the variables $x^{\mu}$,
$p_{\mu}$,
$i$ and $S^{\mu\nu}$:
\BEqA
[\breve x^\mu, \breve J^{\nu \lambda}] &=& \hbar \left(
g^{\mu\nu}
\breve x^\lambda - g^{\mu\lambda}
\breve x^\nu \right),\cr
[\breve p^\mu, \breve J^{\nu \lambda}] &=& \hbar
\left( g^{\mu\nu}
\breve p^\lambda -  g^{\mu \lambda}
\breve p^\nu \right),\cr
[ \breve i, \breve  J^{\mu \nu}] &=& 0 ,\cr
[ \breve S^{\mu\nu}, \breve  J^{\lambda \kappa}] &=& 0.
\EEqA

There is a mock orbital
angular momentum generator of familiar appearance,
\BEq
\label{eq:MOCKORBITAL}
\breve O^{\mu\nu} := -\breve i \left( \breve
x^\mu
\breve p^{\nu} - \breve x^\nu \breve
p^{\mu}\right).
\EEq
$\breve O$ too obeys the Lorent group commutation relations.
We relate $\breve L^{\mu \nu}$ and $\breve
O^{\mu\nu}$ later.

To recover the canonical commutation relations for
$\breve x^{\mu}$ and $\breve p_{\mu}$ we must impose
\BEq
\chi\phi (N-1) \equiv {\hbar\over 2}
\EEq
and assume that
\BEqA
\label{eq:CONTRLIMIT1}
\chi &\rightarrow & 0, \cr
\phi &\rightarrow & 0, \cr
N&\to& \infty .
\EEqA
Then the relations (\ref{eq:EXTERNALCOMMRELSINGLE}) reduce to
the commutation relations (\ref{eq:HA}) of the relativistic
Heisenberg algebra $\3A_H$ as required.

The three parameters $\chi, \phi, 1/N$ are subject to one
constraint $\chi\phi (N-1) = \hbar/2$ leaving two independent
simplifiers. $N$ depends on the scope of the experiment,
and is under the experimenter's control.

We can consider two contractions

1) either $\chi\to 0$
with $N$ constant,

2) or $N\to \infty$
with $\chi$ constant.

They combine into
the  continuum limit $\chi\to 0$, $N\to \infty$.
We  fix one simplifier $\chi$
by supposing that the mass of a probe approaches a finite
limit as
$N\to\infty$.

\subsubsection*{CONDENSATION OF $i$}

Since the usual complex unit $i$ is central and the
simplified $\breve i$ is not, we suppose that the contraction
process includes a projection that restricts the probe to
one of the two-dimensional invariant subspaces of $\breve
i$,  associated with the maximum negative eigenvalue $-1$ of
${\breve i}^2$.  This represents a condensation that aligns
all the mutually commuting hexad spins
$\gamma^{45}(n)$ with each other,  so that
\BEq
\label{eq:CONTRLIMIT}
\gamma^{45}(n)\gamma^{45}(n') \longrightarrow -1,
\EEq
for any $n$ and $n'$.
We call this the {\it condensation of $i$}.

\subsubsection*{ANGULAR MOMENTA}
\label{sec:LORENTZGENERATORS}

As was shown above, three sets of
operators obeying Lorentz-group commutation relations appear
in our theory.
$\breve  L^{\mu\nu}$ represents the simplified {\em orbital angular
momentum}
generators, $\breve S^{\mu\nu}$ represents the {\em spin} angular
momentum, and
$\breve J^{\mu
\nu}$ represents the simplified {\em total angular momentum} generators.
There is a mock orbital angular momentum $\breve O^{\mu\nu}$
(\ref{eq:MOCKORBITAL}).

In this section we show that $\hat O\to \hat L$  in
the contraction limit.

Consider $\breve O^{\mu\nu}$. By definition,
\BEqA
\breve O^{\mu\nu}&=& -\left(\breve
x^\mu \breve p^{\nu} - \breve x^\nu \breve
p^{\mu}\right) \breve i \cr
&=&  +\frac{\chi \phi}{N-1} \left(
\sum_{n=1}^{N-1}
\gamma^{\mu 4}(n)
\sum_{n'=1}^{N-1}
\gamma^{\nu 5}(n') -
\sum_{n=1}^{N-1}
\gamma^{\nu 4}(n)
\sum_{n'=1}^{N-1}
\gamma^{\mu 5}(n') \right) \;
\sum_{m=1}^{N-1}
\gamma^{45}(m) \cr\cr\cr
&=&  +\frac{\chi \phi}{N-1}
\sum_{n}\left(
\gamma^{\mu 4}(n)
\gamma^{\nu 5}(n) -
\gamma^{\nu 4}(n)
\gamma^{\mu 5}(n) \right) \;
\sum_{m}
\gamma^{45}(m) \cr\cr\cr
&& +\frac{\chi \phi}{N-1}
\sum_{n\neq n'}\left(
\gamma^{\mu 4}(n)
\gamma^{\nu 5}(n') -
\gamma^{\nu 4}(n)
\gamma^{\mu 5}(n')
+\gamma^{\mu 4}(n')
\gamma^{\nu 5}(n) -
\gamma^{\nu 4}(n')
\gamma^{\mu 5}(n)\right) \cr\cr
&& \times
( \gamma^{45}(n) + \gamma^{45}(n'))
\cr\cr\cr
&& +\frac{\chi \phi}{N-1}
\sum_{n\neq n'}\left(
\gamma^{\mu 4}(n)
\gamma^{\nu 5}(n') -
\gamma^{\nu 4}(n)
\gamma^{\mu 5}(n')
+\gamma^{\mu 4}(n')
\gamma^{\nu 5}(n) -
\gamma^{\nu 4}(n')
\gamma^{\mu 5}(n)\right) \cr\cr
&& \times
\sum_{m\neq n, m\neq n'}
\gamma^{45}(m) \cr\cr\cr
&=&  -\frac{2\chi \phi}{N-1}
\sum_{n}
\gamma^{\mu \nu}(n)
\gamma^{45}(n) \;
\sum_{m}
\gamma^{45}(m) \cr\cr\cr
&& +\frac{\chi \phi}{N-1}
\sum_{n\neq n'}\left(
\gamma^{\mu 4}(n)
\gamma^{\nu 5}(n') -
\gamma^{\nu 4}(n)
\gamma^{\mu 5}(n')
+\gamma^{\mu 4}(n')
\gamma^{\nu 5}(n) -
\gamma^{\nu 4}(n')
\gamma^{\mu 5}(n)\right) \cr\cr
&& \times
\sum_{m\neq n, m\neq n'}
\gamma^{45}(m) .
\EEqA

Thus, in the contraction limit
(\ref{eq:CONTRLIMIT})-(\ref{eq:CONTRLIMIT1}) when
condensation singles out the eigenspace of
$\gamma^{45}(n)\gamma^{45}(n')$ with
eigenvalue -1,
\BEq
\hat O^{\mu\nu} \longrightarrow \hat
J^{\mu\nu} - \hat S^{\mu \nu} \equiv \hat L^{\mu \nu},
\EEq
as asserted.

\subsection*{SIMPLIFICATION OF THE DIRAC EQUATION}

\subsubsection*{THE DYNAMICS}
\label{sec:DIRACDYNAMICS}

We simplify the Dirac-Heisenberg algebra
$\Ac_{\rm D H}$
to ${\breve \Ac_{\rm DH}} := \Cliff(N{\bf 6_0})$, the
Clifford algebra of a large squadron of cliffordons.

To construct the contraction from ${\breve \Ac_{\rm D H}}$
to $\Ac_{\rm D H}$, we group the generators of $\Cliff(3N,
3N)$ into $N$ hexads $\ga^{\omega}(n)$ ($\omega = 0, \dots,
5$; $n=1,\dots, N$). Each hexad algebra acts on
eight-component real spinors in $\28_0$ (see Appendix E).

{\bf Reminder:} Hexad $N$ is used for the spin of
the quantum. The remaining $N-1$ hexads provide the
space-time variables.

Dirac's spin generators $\hat S^{\mu\nu}$
(\ref{eq:DIRACGENERATORS}) simplify to
the corresponding 16 components of the tensor
\BEq
\label{eq:DIRACSPINOPERATOR}
\breve S^{\omega \rho}:= \frac{\hbar}{2}\,
\gamma^{\omega \rho}(N),
\EEq
where $\omega , \rho = 0, \dots , 5$
and $\mu,\nu = 0, \dots, 3$.

Then {\bf the most natural choice for Dirac's dynamics} is
\BEq
\label{DYNAMICS}
\tilde{D} := \frac{2\phi}{\hbar^2} \; \breve
S^{\omega\rho}
     \breve L_{\omega \rho},
\EEq
where ({\it cf.} (\ref{eq:ANGULARMOMENTA}))
\BEq
\breve L_{\omega \rho}:= \frac{\hbar}{2}\sum_{n=1}^{N-1}
\gamma_{\omega \rho}(n),
\EEq
and
\[
\frac{2\phi}{\hbar^2}=\frac{1}{\hbar \chi (N-1)}.
\]

The proposed dynamical operator is invariant under
$\SO(3,3)$. This symmetry group incorporates
and extends the
$\SO(3, 2)$ symmetry possessed by Dirac's dynamics for an
electron in de-Sitter space-time
\cite{DIRAC35}:
$$
D'=\frac{1}{\hbar R}{\hat S}^{\omega \rho} {\hat
O}_{\omega\rho}
- mc,
$$
where ${\hat S}^{\omega \rho}$ and $\hat O_{\omega\rho}$ are
the five-dimensional spinorial and orbital angular momentum
generators and $R$ is the radius of the de-Sitter universe.
That group
unifies translations, rotations and boosts,
but not the $i$.

\subsubsection*{REDUCTION TO THE POINCAR\'E GROUP}

We now assume a condensation that reduces $\SO(3,3)$ into
$\SO(1,3)\x\SO(2)$. Relative to this reduction, the $\tilde
D$ of (\ref{DYNAMICS}) breaks up into
\BEqA
\tilde{D} &=& \frac{\phi}{2} \, \gamma^{\omega\rho}(N)
\sum_{n} \gamma_{\omega \rho}(n) \quad \quad \quad
\quad
\quad \quad
(\omega, \rho = 0, 1, \dots , 5 ) \cr\cr\cr &=& \phi \,
\gamma^{\mu 5}(N)
\sum_{n} \gamma_{\mu 5}(n) + \phi \,
\gamma^{\mu 4}(N)
\sum_{n} \gamma_{\mu 4}(n) + \phi \,
\gamma^{\mu\nu}(N)
\sum_{n} \gamma_{\mu\nu}(n) \cr
&&+ \, \phi \, \gamma^{45}(N)
\sum_{n} \gamma_{45}(n) \cr\cr\cr
&=&
\gamma^{\mu 5} \, {\breve p}_\mu - \frac{\phi}{\chi} \,
\gamma^{\mu 4} \, {\breve x}_{\mu} +
\frac{2\phi}{\hbar}\,
\gamma^{\mu\nu} \, {\breve L}_{\mu\nu} + (N-1) \phi \,
\gamma^{45} \, {\breve i}.
\EEqA

In the condensate  all the operators $\gamma^{45}(n)
\gamma_{45}(n')$ attain their minimum eigenvalue $-1$.
Then
\BEq
     (N-1) \, \phi \,
\gamma^{45}\, {\breve i} \; \longrightarrow
\;- \frac{\hbar}{2 \chi}.
\EEq
and the
dynamics becomes
\BEq
\label{eq:DIRACDYNAMICSCONTRACTED}
\tilde{D} \; = \;
\gamma^{\mu5} \, {\breve p}_\mu -
\frac{\phi}{\chi}\,
\gamma^{\mu 4} \, {\breve x}_{\mu} +
\frac{2\phi}{\hbar}\,
\gamma^{\mu\nu} \, {\breve L}_{\mu\nu} - m_\chi c,
\EEq
with rest mass
\BEq
\label{eq:ADJUSTEDMASS}
m_\chi = \frac{\hbar}{2 \chi c}.
\EEq

Here we identify the usual Dirac gammas $\ga^{\mu}$ for $\mu=0,
\dots, 3$ of $\Cliff(\Mbb )$ with second-degree elements
of the last hexad:
\BEq
\gamma^{\mu} \cong \tilde \gamma^{\mu} := \gamma^{\mu 5} (N)
\EEq

For sufficiently large $N$ this $\tilde{D}$ reduces to the usual
Dirac dynamics.

We identify the  mass $m_{\chi}$
with the $N$-independent mass
$m$ of the Dirac equation
for the most massive individual quanta
that the condensate can propagate without melt-down,
on the order of the top quark:
\BEq
m_{\chi}\sim 10^2 \GeV,\quad \chi\sim 10^{-25} \sec .
\EEq
The universe is $\sim 10^{10}$ years old.
This leads to an upper bound
\BEq
N_{\rm Max} \sim 10^{41}.
\EEq

Here $\chi$ is independent of $N$
as $N\to \infty$ and that $\phi \sim 1/N \to 0$
as $N\to \infty$ even for finite $\chi$.

In experiments near
the Higgs energy, $p\sim \hbar /\chi$.  If we also determine
$N$ by setting $x\sim N\chi$ then all  four terms in
(\ref{eq:DIRACDYNAMICSCONTRACTED})  are of the same order of
magnitude.

To estimate experimental effects, however,
we must take gauge transformations into account.
These transform the second term away in the continuum limit.
This refinement of the theory is still in progress.

\newpage

\section*{ \center CONCLUSIONS}
\addcontentsline{toc}{section}{CONCLUSIONS}

     Like classical Newtonian mechanics, the Dirac equation has
a compound (non-semisimple) invariance group. Its variables
break up into three mutually commuting sets: the
space-time-energy-momentum variables ($x^{\mu}$,
$p_{\mu}$), the spin variables $\ga^{\mu}$,
and the imaginary unit $i$.

To unify them we replace the space-time continuum
by an aggregate of $M < \infty$ finite elements,
chronons,
described by spinors with $\sim 2^{M/2}$ components.
Chronons have Clifford-Wilczek statistics,
whose simple operator algebra
is generated by  units $\ga^m$, $m = 1, \dots, M$.
We express
all the variables $x^{\mu}, p_{\mu}, \gamma^{\mu}$ and $i$
as polynomials in the $\ga^m$.
We group the $M=6N$ chronons into $N$ hexads for this
purpose, corresponding to tangent spaces;
the hexad is
the least cell that
suffices for this simplification.
There are three simplifiers
$\chi,
\phi, 1/N$, all approaching 0 in the continuum limit,
     subject to the constraint $\chi \phi \, (N-1) = \hbar/2$ for
all
$N$.

In the continuum limit the Dirac mass becomes infinite.
In our theory, the finite Dirac
masses in nature are consequences of
a finite atomistic quantum space-time structure with
$\chi>0$.

The theory predicts a certain spin-orbit coupling
$\gamma^{\mu\nu} L_{\mu\nu}$ not found in the Standard Model,
and vanishing only in the continuum limit. The experimental
observation of this spin-orbit coupling would further indicate the
existence of a chronon.

In this theory,
the spin we see in nature
is a manifestation of
the (Clifford) statistics of
atomic elements of space-time,
as Brownian motion is of the atomic elements
of matter.
As we improve our theory
we will interpret better other indications of
chronon structure that we already have,
and
as we improve our measuring techniques
we shall meet more such signs.

\newpage

\section*{ \center  APPENDIX A \\ \center OPERATIONALITY AND
THE GEOMETRICAL NATURE OF PHYSICS}

\addcontentsline{toc}{section}{APPENDIX A}



A good physical theory
should be based on the following

\paragraph{ Operational Postulate:} If we do so-and-so, we
will find such-and-such.

\vskip15pt

Generalizing these ``doing and finding'' actions (carried
out by an experimenter) to the operations going on in Nature
independent of any observer, and taking them as the basic
units of our theory, we arrive at systems analogous to
the ones frequently used in mathematics. By applying to
these systems the rules of mathematical
inference we derive new predictions that could be
tested in experiment.

One of the most remarkable mathematical systems often
used in physics is the system of geometry in which the
basic structural element of paramount importance is an
incidence basis \cite{mihalek}. The incidence basis
consists of two (incidence) classes, disjoint or not,
often called the class of points $\Pc$ and the
class of lines
$\Lc$, and the incidence relation, usually denoted by
$\circ$, between the elements of the two classes.

     Traditionally the notion of incidence
involves a definite rule of pairing of elements from
$\Pc$ with elements from
$\Lc$, or, more specifically, their ordered pairing.

To realize the ordered pairing, the notion of
an ordered pair $(a,\,b)$ is introduced, where $a$ and
$b$ are the elements of the pair with $a$ the first and $b$
the second. The set of all ordered pairs $(a, \, b)$, $a \in
\Pc$, $ b
\in \Lc$ is called the Cartesian product of $\Pc$ and $\Lc$
and designated by $\Pc \times \Lc$.

The study of incidence, then, involves certain subset
of $\Pc \times \Lc$ whose specification establishes an
incidence relation between the elements of the two classes,
in analogy to how it is done, for example, in elementary
Euclidean geometry.

The properties of a particular incidence
relation are called axioms of the corresponding
incidence basis. Together they define a
geometry.

It is a remarkable fact that both
most important theories of contemporary physics, quantum
theory and relativity, can be regarded as geometries with
some definite forms of the incidence bases. Moreover, any
physical theory obeying The Operational Postulate must have
the form
     similar to some (generalized) geometry.
     This is because the postulate itself has
the incidence structure: our ``doings'' and
``findings'' can be regarded as the elements
of some incidence classes, and the correlation between the
two may be viewed as
an incidence relation.

Formulating all of physics in the
form of an incidence basis is a tedious task.
It involves considerable labor of stating and proving
the theorems of the corresponding geometrical system and
relating them back to the experience in the form of some
definite operational procedures.

What is more important, however, is that trying to fit
everything in the usual geometrical framework,
would eventually lead to the same limitations that were
encountered in the classical axiomatic method.
This can be seen as follows:

     As we know, one of the main
problems of the classical axiomatic method is the ``proof''
of internal consistency of a given geometry, so that no
mutually contradictory theorems can be deduced from the
axioms of an incidence basis under consideration.

A general method of
``solving'' this problem is based on exhibiting a basis,
called a consistency basis (a model or
interpretation), so that every axiom of the original basis
is converted into a ``true'' statement about the model
\cite{nagel, mihalek}. For example, it is
possible to show that the model for the
Euclidean geometry can be grounded on the
axiomatic system of elementary arithmetics,
etc. Unfortunately in most cases, including this one, the
method just shifts the problem of consistency of a geometry
to the  problem of consistency of the model itself. If the
geometry is such that a finite model can be built
whose incidence classes contain a finite number of elements,
we may try to establish the consistency of the geometry by
direct inspection of the model and determining whether its
elements satisfy the original axioms. However for most of
the axiomatic systems that are important in mathematics and
used by physicists finite consistency bases cannot be
constructed.

In this respect, we may regard a physical theory obeying
The Operational Postulate as a
consistency basis (a model, interpretation) of a geometry we
choose to work with. Although from this perspective, of
course, everything looks turned around --- usually the
theory, not the results of experiments, is regarded as a
model --- a closer look reveals that this is just an
expression of our constant desire to use mathematics and its
methods in discovering the workings of the world around us.
Here the model is, in a sense, richer than the corresponding
geometry,  exactly how it is often in mathematics. While in
mathematics the extra properties of the model might obcure
the derivation of interesting theorems, in physics these are
the assets, and their discovery constitutes the ultimate
goal and purpose of science.

An alternative to the consistency basis method is
the method of an absolute proof of consistency in
which the consistency of the system is sought to be
established without assuming the consistency of some other
model system
\cite{nagel}. As far as physics is concerned this method is
important in the mathematical part of the theory ---
deriving the predictions (see below) --- as providing
justification for the use of ordinary logic in manipulating
the recorded results of experiments. However even this
ambitious method has failed to solve the general problem of
consistency. As G\"{o}del showed, it is impossible in
principle to establish the internal logical consistency of a
large class of axiomatic systems --- including elementary
arithmetics --- without adopting principles of reasoning so
complicated that their own internal consistency is no less
doubtful than that of the systems themselves.

Another important limitation of the classical axiomatic
method discovered by  G\"{o}del is that any axiomatic
system within which arithmetic can be developed is
essentially incomplete, meaning that given any
consistent set of axioms, there are true arithmetical
statements that cannot be derived from that set.

It is of no surprise then that the chain

\vskip5pt

\noindent do experiment $\rightarrow$ generalize
$\rightarrow$ idealize $\rightarrow$ axiomatize (and
possibly modify by introducing additional elements to the
incidence basis)
$\rightarrow$ derive
$\rightarrow$ translate into operational procedure
$\rightarrow$ check by doing experiment,

\vskip5pt

\noindent would miss or distort some important physical
content and introduce some inconsistencies of the
geometry that will result in the
impossibility of performing the suggested experiments, or
simply lead to unverifiable predictions.

   All
theories of physics suffer in one way or
another from the above mentioned limitations of the very
axiomatic method. New
phenomena may always be discovered whose existence cannot be
predicted (or disproved) within the axiomatic system, no
matter how full the incidence relation of
The Operational Postulate of Physics we choose. That is why
the whole body of physics must be formulated in the
operational terms. If then it turns out possible to
construct an axiomatic system that helps us make new
predictions, good --- we accept it as a working theory. If
not, we may be led to a new scheme (based on a different from
axiomatic method) in terms of which physics would be
formulated.



So far the usual mathematics with its axiomatic method
has worked remarkably well. In particular, the system of
geometry on which quantum mechanics is based
has been very successful within its domain of applicability.
Its basic ``ingredients'' have a well
defined operational meaning. By itself quantum
mechanics does not require any interpretation. It is its own
interpretation.

Here, following Finkelstein
\cite{DRF96, DRF99}, we briefly summarize the quantum
mechanical operational terminology which in this work
will be used throughout.

We start with the kets, which represent sharp initial actions
on the system under study. These initial actions are
typically of the form: ''release from the source and
then select with a filter''.

Kets do not represent states of the system,
contrary to popular belief. Kets represent what (kind of
filtering) we do to the system in the beginning of each
experiment. The notion of a state does not make much sense in
quantum mechanics, especially if applied to one individual
quantum in experiment.

Similarly, and dually, the bras represent final filtering
actions followed by detection with
an appropriate counter. Also, operators
represent all possible operations on the system. The kets
and bras are special kinds of operators. Mathematically they
can be regarded as the elements of the minimal left and right
ideals of system's operator algebra.

If we do a polarization experiment with a photon,

\vskip10pt

       (source $\rightarrow$ initial polarizer) $\rightarrow$
time evolution $\rightarrow$ (final polarizer $\rightarrow$
detector),

\vskip10pt

\noindent no matter what we do to the photon after the
initial action is completed,  we will never be able to tell
by what filter (vertical, horizontal, circular) it was
selected initially. To find that out we would have to go
back to the initial filter and look at it.

If the photon had a state, by ``determining'' it we would be
able to tell unambiguously what initial filter had been used
and what final filter will have to be used in order for the
experiment to end up with a counter click. Such
determination is possible in classical mechanics where the
notion of the state is meaningful, but not in quantum
mechanics.

It turns out that the superposition principle (which is a
typical reason for retaining the non-operational ``state''
terminology in quantum mechanics) can be naturally
formulated for the initial actions. The initial actions can
be viewed as the elements of a Hilbert space, so all the
usual mathematical formalism of quantum mechanics survives.

     Thus, we again start with two spaces, the ket space $V$ of
initial selective actions on the system and its dual
space $V\adj$ of the final selective actions.

We have an adjoint that maps the two (see
Appendix B). If the adjoint is positive definite
we get the usual quantum mechanics with the positive definite
metric.

We contract an initial ket with a final bra using that
adjoint, to get the transition amplitude for the
two-stage experiment of the form:

\vskip10pt

       (source $\rightarrow$ initial selective act) $\rightarrow$
(final selective act $\rightarrow$
detector).

\vskip10pt

{\bf The operational meaning of the adjoint} is the
following:

The final bra $\bra \psi |=|\psi \ket\adj$ which is the
adjoint of an initial ket $|\psi \ket$, is such a final act
that the transition $\bra \psi | \leftarrow |\psi \ket$ is
always allowed (every time we send a quantum it goes
through both initial and final filters and the detector
clicks). Problems might appear with an indefinite adjoint. In
that case some transitions $\bra \psi | \leftarrow |\psi
\ket$ never happen because sometimes $\bra \psi |\psi
\ket =0$.

     Using the adjoint we could naturally introduce metric on
both $V$ and $V\adj$, but its operational
interpretation would be obscure. It is always better to
keep the operational difference between initial (ket) and
final (bra) spaces in mind and talk about the adjoint, not
the metric

     Classical mechanics can be easily cast into similar
operational form by switching off superposition of actions.

And finally, {\bf The superposition principle}:

An initial act $|\psi_3\ket$ is a coherent superposition of
initial acts $|\psi_1\ket$ and $|\psi_2\ket$,
\[
|\psi_3\ket = |\psi_1\ket + |\psi_2\ket,
\]
if  every final act $\bra \phi |$ that occludes $|\psi_1\ket$
and $|\psi_2\ket$ also occludes $|\psi_3\ket$.
In other words, if the transitions
$\bra \phi | \leftarrow |\psi_1\ket$ and $\bra \phi |
\leftarrow |\psi_2\ket$ never happen,
then the transition $\bra \phi | \leftarrow |\psi_3\ket$
(for the same actions $\bra \phi | \in V\adj$) never happens
either. In Dirac's notation,
\[
(\bra \phi |\psi_1\ket =0 \; {\rm AND} \; \bra \phi
|\psi_2\ket =0) \Longrightarrow (\bra \phi
|\psi_3\ket =0).
\]

And dually for the superposition of final actions.

\vskip10pt

\paragraph{EXAMPLE.} Spin-1 particle in Stern-Gerlach
experiment.

\noindent Here,
\[
|\psi_1\ket =| +1 \ket, |\psi_2\ket=| -1 \ket, \bra \phi
|=\bra 0 |,
\]
\[
|\psi_3\ket = |\psi_1\ket + |\psi_2\ket =
|+1 \ket + | -1 \ket,
\]
(both filters are opened),
with
\[
\bra 0 |+1 \ket =0, \quad \bra 0 |-1 \ket =0 \quad \quad
({\rm transitions \; never \; happen)}.
\]
Then
\[
\bra 0 |\psi_3\ket =0,
\]
     meaning that this transition never happens either.
(Of course here all the vectors in question must be
properly normalized.)

\vskip10pt

Thus in quantum mechanics we don't need states.
We don't need nouns (what system {\it is}), as Finkelstein
puts it, all we need are verbs (what we {\it do} to the
system). In fact we {\it define} a particular system by what
we can {\it do} to it.

By inventing non-operational concepts like ``state
vector'', ``collapse of the wave function'', etc., it is easy
to drive ourselves into many contradictions and
paradoxes. Keeping the operational meaning of quantum
mechanics in mind, however, can help us avoid such pitfalls.

\newpage

\section*{ \center APPENDIX B \\ \center ADJOINT OPERATOR}
\label{sec:ADJOINT}

\addcontentsline{toc}{section}{APPENDIX B}

\subsection*{DEFINITION}

Let $\Fbb$ be a field of real ($\Rbb$) or complex
($\Cbb$) numbers and $V$ be a  {\it right}
$\Fbb$-linear space with {\it some arbitrary} basis $\{e_a \,
|
\, a=1,
\,
\dots, \, {\rm dim}V \}$ in it.

An operator $\dag$,
\BEqA
\dag : \; V \longrightarrow V^\dag, \; \psi = e_a \psi^a
\equiv (\psi^a) \;
\mapsto \; \dag \psi := \psi\adj = \psi\adj_a e^a \equiv
(\psi\adj_a) ,
\EEqA
from $V$ to its {\it left} $\Fbb$-linear dual space
$V^\dag$ (with {\it some} basis $\{e^a \, | \, a=1, \,
\dots, \, {\rm dim}V \}$) is said to be

\noindent
$\circ$ {\it singular} when
\BEq
\exists \; \psi \neq 0 \, : \, \psi\adj = 0;
\EEq

\noindent
$\circ$ {\it symmetric} when
\BEq
\phi\adj \cdot \psi = \psi\adj \cdot \phi;
\EEq

\noindent
$\circ$ {\it Hermitian-symmetric} when
\BEq
\phi\adj \cdot \psi = (\psi\adj \cdot \phi)^{C};
\EEq

\noindent
$\circ$ {\it definite} when
\BEq
\forall \; \psi \neq 0, \; \psi\adj \cdot \psi \neq 0;
\EEq

\noindent
$\circ$ {\it positive (definite)} when
\BEq
\forall \; \psi \neq 0, \; \psi\adj \cdot \psi > 0.
\EEq
Here
\BEq
\phi\adj \cdot \psi := \phi\adj_a \, \psi^a \equiv \bra \phi
| \psi \ket
\EEq
means the {\it contraction} of
$\phi\adj \in V^\dag$ with $\psi \in V$ and
$^C$ stands for the complex conjugation.

By definition, an {\it adjoint operator} is a mapping
$\dag : V \longrightarrow V^\dag$ that is
antilinear, non-singular and Hermitian-symmetric.
(When $\Fbb = \Rbb$, an adjoint operator is linear
and symmetric.)

Defining
\BEq
\dag e_a = e_a\adj := M_{ba}e^b,
\EEq
we get
\BEqA
\psi\adj = \dag (e_a \psi^a) := \psi^{aC} (\dag e_a)
= M_{ba} \psi^{aC} \, e^b,
\EEqA
leading to
\BEq
\psi\adj_b =  M_{ba} \psi^{aC}.
\EEq

The contraction of $\psi\adj \in V\adj$ with $\phi \in V$ is
then
\BEq
\psi\adj \cdot \phi =  M_{ba} \psi^{aC} \phi^b \equiv
(M \psi^{C})^T
\phi,
\EEq
$^T$ meaning transposition.

The assumption of Hermitian symmetry of $\dag$ implies
\BEq
M_{ba} = M^C_{ab},
\EEq
or, equivalently,
\BEq
M = M^{CT} =: M^*.
\EEq

The matrix $M_{ab}$ of $\dag$ (called the {\it
transition metric}) is defined relative to some {\it
arbitrary} bases of the corresponding spaces $V$ and
$V\adj$.

{\it In general}, it is {\it not} true that
$e_a\adj = e^{a}$. Rather, in physical applications the form
of the transition metric is decided {\it operationally},
namely, relative to the properly defined actions belonging
to some initial and final frames. For example, in the {\it
standard non-relativistic} quantum mechanics it is often
possible to find the frames relative to which $M_{ab}$ is
the identity matrix ($\dag$ is positive definite) and
$e_a\adj = e^{a}$, so that $\psi\adj =
(\psi\adj_a)$ is determined by the "usual" rule
"transpose + complex-conjugate". However, in the more
complicated situations $\dag$ may be indefinite.

Because $\dag$ is non-singular, its (antilinear) inverse
$\dag^{-1}: \; V\adj \longrightarrow V$ can be defined by
the condition
\BEq
\dag^{-1}: \psi\adj \mapsto \dag^{-1} \psi\adj := \psi.
\EEq
If we give the action of $\dag^{-1}$ on the basis elements
of
$V\adj$,
\BEq
\dag^{-1} e_b := e_f M^{fb},
\EEq
then
\BEqA
\dag^{-1} \psi\adj &=& \dag^{-1}(\psi\adj_b \, e^b) :=
(\dag^{-1} e^b) \, {\psi\adj_b}^C \nonumber \\
&=& (\dag^{-1} e^b) \, ( M_{ba} \psi^{aC})^C \nonumber \\
&=& (\dag^{-1} e^b) \, {M_{ba}}^C \, \psi^a \nonumber \\
&=& e_f M^{fb} \, M_{ba} \psi^a \nonumber \\
&:=& e_a \, \psi^a,
\EEqA
with
\BEq
M^{fb}M_{ba} := \delta^f_a.
\EEq

Usually, $\dag^{-1}$ is denoted by the same symbol $\dag$,
regarding an adjoint operation as involutory anti-automorphism
of the action semigroup.

Now, any {\it linear} operator $A: \; V
\stackrel{\rm linear}{\longrightarrow} V$ acting on $V$ can
be represented (relative to some basis $\{ e_a \}$) by its
matrix $A^b_a$:
\BEq
\label{eq:MATRIXOFOPERATOR}
A : e_a \mapsto A e_a := e_b \, A^b_a.
\EEq
Then,
\BEq
\psi \mapsto A \psi = A (e_a \psi^a)
:= e_b \, A^b_a \, \psi^a.
\EEq

$A$ can be also considered as acting (linearly)
on the elements $\psi\adj$ of the dual space $V\adj$ by the
rule
\BEqA
\psi\adj \mapsto \psi\adj A &=& (\psi\adj_a \, e^a) A
\nonumber \\
&:=& \psi\adj_a \, (e^a A) \nonumber \\
&:=& \psi\adj_a \, A^a_b \, e^b.
\EEqA

This allows us to write $A$ in the form
\BEq
\label{eq:ANOTHERDEFOFOPERATOR}
A = e_b \otimes A^b_a \, e^a ,
\EEq
with $e_b$ and $e^a$ {\it now} being the elements of two
{\it mutually dual} bases
\BEq
\label{eq:DB}
e^a \cdot e_b := \delta^a_b.
\EEq
Acting with
({\ref{eq:ANOTHERDEFOFOPERATOR}) on the basis vector $e_a$
recovers (\ref{eq:MATRIXOFOPERATOR}).

This form of writing allows us to consider
$A$ as acting on {\it both} dual spaces, $V$ and $V\adj$, by
the usual rule: vectors are being acted upon from the
left and dual vectors from the right.

This also gives the {\it
resolution of the identity} (in the {\it dual bases}),
\BEq
{\bf 1} =  \sum_{a} e_a \otimes e^a.
\EEq
Note that when $e_a\adj = e^a$ we get the usual
resolution of the identity of the standard quantum mechanics,
\BEq
{\bf 1} =  \sum_{a} e_a \otimes e_a\adj.
\EEq

We may define the {\it adjoint of $A$}, denoted by
$A\adj : V \longrightarrow V$, as
\BEq
\label{eq:ADJOINTOFOPERATOR}
\psi\adj \cdot A\adj \phi := ( \phi\adj \cdot A \psi
)^\Crm \;\;\;\;\;  (\forall \, \psi, \, \phi \, \in V).
\EEq

Direct calculation leads to
\BEq
{A\adj}^a_b = M^{af} {A^{CT}}^f_g M_{gb},
\EEq
or
\BEq
A\adj = M^{-1} A^{*} M.
\EEq
This formula is valid with respect to arbitrary bases of
$V$ and $V\adj$.

We also define the {\it unitary} operator $U$ acting on
the vector space $V$ (or $V\adj$) as an operator obeying
\BEq
U\adj U = U U\adj = {\bf 1},
\EEq
leading to the condition
\BEq
(U\psi)\adj \cdot U \phi = \psi\adj \cdot
\phi.
\EEq

The preceding consideration can be generalized from
vector spaces to {\it modules}. Instead of the operator
$\Crm$ acting on the complex field $\Cbb$ we must now
consider an appropriate {\it anti-automorphism} of the
corresponding {\it ring} over which the module is built.
As an example of such generalization we can mention a module
over the division ring of quaternions with the usual
quaternion conjugation.

\subsection*{THE VAN DER WAERDEN NOTATION}

In spinorial
general relativity the notion of adjoint is
generalized to include the anti-linear mappings.
     For completeness we give a brief summary
of the special {\it dot notation} developed for such cases by
Van der Waerden.

Generalizing from the symmetric case, an {\it adjoint
operator} (symmetric or antisymmetric) is a mapping
\BEqA
\Cb : \; V \longrightarrow \Vd, \; \psi = e_a \psi^a
\equiv (\psi^a) \;
\mapsto \; \Cb \psi := \psid = \psi_\ad e^\ad \equiv
(\psi_\ad) ,
\EEqA
which is
{\it anti}linear, non-singular and Hermitian-(anti)symmetric.
(When $\Fbb = \Rbb$, an adjoint operator is linear
and (anti)symmetric.)

Defining
\BEq
\Cb e_a \equiv \ed_a := C_{\bd a}e^\bd,
\EEq
we get
\BEq
\psid = \Cb (e_a \psi^a)
= {\bar \psi}^{a} (\Cb e_a)
:=  C_{\bd a} {\bar \psi}^{a} \, e^\bd,
\EEq
leading to
\BEq
\psi_\bd := C_{\bd a} {\bar \psi}^{a} \quad .
\EEq

The contraction of $\psid \in \Vd$ with $\phi \in V$ is
then
\BEqA
\psid \cdot \phi &=&
     C_{\bd a} {\bar \psi}^{a} \; ( e^\bd \cdot e_f ) \;
\phi^f =  C_{\bd a} {\bar \psi}^{a} \delta^\bd_f \, \phi^f
\cr &\equiv& C_{fa} \, {\bar \psi}^{a} \phi^f .
\EEqA
Here,
\BEq
( e^\bd \cdot e_f ) := \delta^\bd_f \quad ,
\EEq
and
\BEq
C_{fa} := C_{\bd a} \delta^\bd_f.
\EEq

The assumption of Hermitian (anti)symmetry of $\Cb$ implies
\BEq
C_{fa} = (-){\bar C}_{af}
\quad .
\EEq

Because $\Cb$ is non-singular, its (antilinear) inverse
$\Cb^{-1}: \; \Vd \longrightarrow V$ can be defined by
the condition
\BEq
\Cb^{-1}: \psid \mapsto \Cb^{-1} \psid := \psi.
\EEq
We have
\BEqA
\psi &=& \Cb^{-1} \psid \cr
&=& \Cb^{-1} (\psi_\ad e^\ad) \cr
&=&  (\Cb^{-1} e^\ad) \, {\bar \psi_\ad} \cr
&=& e_b \, C^{b \ad } {\bar \psi_\ad},
\EEqA
where we have defined
\BEq
\Cb^{-1} e^\ad := e_b \, C^{b \ad }.
\EEq
Correspondingly,
\BEq
\psi^b = C^{b \ad } {\bar \psi_\ad}.
\EEq

We also have
\BEq
     C^{ b \ad} C_{\ad f} = \delta^b_f \, , \quad C_{\ad f}C^{ f
\bd} =
\delta^\ad_\bd \quad .
\EEq

In the spinor algebra of special relativity developed in
\cite{RUMER},
$V= \Cbb^2$, and there defined the
{\it main antisymmetric bilinear form} on $V= \Cbb^2$ by
\BEq
G(\psi, \phi) := (\Cb \psi | \phi) \equiv (\psid |\phi)
= {\bar C_{fa}} \, \psi^a \phi^f \;.
\EEq
Similarly, on $\Vd=\dot{\Cbb}^2$,
\BEq
\Gd(\psid, \phid) := (\psid | \Cb^{-1} \phid)^* \equiv
(\psid |\phi)^* = {\bar C^{\ad \fd }} \, \psi_\ad \phi_\fd
\;.
\EEq
Thus, $C_{ab}$ is antisymmetric and can be written as
\BEq
(C_{fa}) := \left[\matrix{ 0 & 1 \cr - 1 & 0 }\right].
\EEq
Also,
\BEq
(C^{\fd\ad}) := \left[\matrix{ 0 & -1 \cr 1 & 0 }\right].
\EEq

\newpage

\section*{\center  APPENDIX C\\ \center IDEALS}
\label{sec:IDEALS}

\addcontentsline{toc}{section}{APPENDIX C}

Very often an algebra can be constructed
from smaller algebras by some rules of assembling
them (see, for example, \cite{BENNTUCKER}, p. 317). One
particularly relevant to our theory example of such
construction is the {\it direct sum} of two algebras, when
an algebra $\Ac = \Bc
\oplus \Cc$, as a vector space, is a direct sum of the
vector spaces $\Bc$ and $\Cc$, and
$\Bc
\bullet
\Cc = \Cc \bullet \Bc = 0$, where $\bullet$ indicates the
product on $\Ac$, be it
associative, commutator (Lie), or else.\footnote{For
associative algebras we write $\Bc
\Cc = \Cc \Bc = 0$; for Lie algebras we write $[ \Bc ,
\Cc ] = 0$.} An
algebra that can be written as a direct sum of several
algebras is called {\it reducible}.

Reducible algebras contain invariant subalgebras, also knows
as ideals. There are different kinds of ideals: left, right,
or two-sided.

A {\it left ideal} is a {\it subspace}
$\Ic \subset \Ac$ such that $\Ac \bullet \Ic
\subset \Ic$.

A {\it right ideal} is a {\it subspace}
$\Ic \subset \Ac$ such that $\Ic \bullet \Ac
\subset \Ic$.

A {\it two-sided ideal} is a {\it subspace}
$\Ic \subset \Ac$ such that $\Ac \bullet \Ic \bullet \Ac
\subset \Ic$.

It is clear that all these ideals are also {\it subalgebras}
of $\Ac$. Moreover, if $\Ac = \Bc \oplus \Cc$ then {\it both}
$\Bc$ and $\Cc$ are its {\it two-sided}
ideals.\footnote{Proof:
\BEqA
(\Ac \bullet \Bc )\bullet \Ac &=& ((\Bc \oplus \Cc)\bullet
\Bc)
\bullet (\Bc
\oplus
\Cc) \nonumber \\
     &=& ((\Bc \bullet \Bc ) \oplus ( \Cc \bullet \Bc ))
\bullet (\Bc
\oplus
\Cc) \nonumber \\
&=&  (\Bc \bullet \Bc ) \bullet (\Bc \oplus
\Cc)
\nonumber \\ &=& ((\Bc \bullet \Bc) \bullet \Bc)\oplus ((\Bc
\bullet \Bc)
\bullet \Cc)) \subset \Bc \quad .
\EEqA
} As to the one-sided ideals, it is
possible that for example
$\Bc$ is such an ideal, but $\Cc$ is not.\footnote{The
relativistic Heisenberg algebra (\ref{eq:HA}) is an example,
where $i$ generates a one-sided ideal, but $\hat p$
and $\hat x$ do not.}

Let us suppose now that $\Ac = \Bc + \Cc$ as a vector space,
and let us define an equivalence relation in $\Ac$ by
$a \sim b$ if $a=b+c$ where $c \in \Cc$. Denote the
equivalence class of $a$ in the usual way by $[a]$.

The equivalence classes so defined can be made into a vector
space with the rules of addition and scalar multiplication
as follows:
\BEq
[a]+[b]=[a+b],
\EEq
and
\BEq
\lambda [a] = [\lambda a].
\EEq

The question arises if these equivalence classes can also be
made into an algebra. An obvious choice for multiplication
is
\BEq
[a][b]=[a \bullet b].
\EEq
However, if $c_1,c_2\in\Cc$
then
\BEqA
[a][b]&=&[a+c_1][b+c_2]\nonumber \\
&=& [(a+c_1)\bullet(b+c_2)] \nonumber \\
&=& [a \bullet b + a \bullet c_2 + c_1 \bullet b + c_1
\bullet c_2] \nonumber \\
&=& [a \bullet b + (a \bullet c_2 + c_1 \bullet b + c_1
\bullet c_2)] ,
\EEqA
which means that $(a \bullet c_2 + c_1 \bullet b + c_1
\bullet c_2)\in \Cc$. This is true if, and only if,
$\Cc$ is a {\it two-sided} ideal of $\Ac$. The algebra of
equivalent classes so constructed is called {\it
quotient algebra} of
$\Ac$ modulo $\Cc$, which is denoted by $\Ac/\Cc$.

     This is
precisely the way in which many important algebras
(including the tensor, the Clifford, etc., algebras) are
defined in modern mathematics. We also use this method to
define various forms of quantification.

\newpage

\section*{\center  APPENDIX D \\ \center LOCALIZATION
PROBLEM}
\label{sec:LOCALIZATION}

\addcontentsline{toc}{section}{APPENDIX D}

\subsection*{THE PLANK LIMIT: MEASURING A FIELD AT A POINT}

Gravity and quantum theory provide a well-known
qualitative lower bound to the size ($\sim \tau_{\rm p}$) of
a space-time cell over which an average field of any kind can
be measured with arbitrary precision.

{\bf Argument:}

STEP 1: Localization of the field meter (say, a test
particle) within a space-time cell $\sim (c\tau_{\rm p})^4$.
By uncertainty relation,
\[
\Delta \varepsilon \; \tau_{\rm p} = \hbar.
\]

STEP 2: Schwarzschild radius associated with
$\Delta \varepsilon$ is
\[
R_{\rm Sch}=\frac{2G\, \Delta M}{c^2} = \frac{2G\,
\Delta \varepsilon}{c^4} =
\frac{2 G \, \hbar}{c^4 \; \tau_{\rm p}}.
\]

STEP 3: To avoid formation of the horizon, set
\[
R_{\rm Sch} < c \, \tau_{\rm p}.
\]

STEP 4: This leads to the Plank limit,
\BEq
\tau_{\rm p} > \sqrt{\frac{2 G \, \hbar}{c^5}} = 7.6\times
10^{-44} \; {\rm s},
\EEq
which corresponds to the Plank energy
\[
\varepsilon_{\rm p} = \hbar / \tau_{\rm p} =  1.4 \times
10^{9} \; {\rm J} = 8.7 \times
10^{18} \; {\rm GeV},
\]

Here,

$c = 2.998 \times 10^{8} \; {\rm m}/{\rm s}$

$G = 6.673 \times 10^{-11} \;
{\rm m}^3/{\rm kg}\; {\rm s}^2$

$\hbar = 1.054 \times 10^{-34} \;
{\rm J}\,{\rm s}$

1 GeV = $1.602 \times 10^{-10}$ J = $1.783 \times 10^{-27}$
kg

1 GeV$^{-1}$ = $1.973 \times 10^{-14}$ cm = $0.1973 $ fm

\subsection*{THE MANY-CELL HORIZON LIMIT: MEASURING A FIELD
OVER A REGION OF SPACE-TIME}

The validity of QFT requires not only that the field at one
space-time point can be measured, but also that the field at
{\it every} point of an experimental region at one
time-instant be measured!

{\bf Argument:}

STEP 1: Localization of the field meter (say, a test
particle) within a space-time cell $\sim (c\tau)^4$. By
uncertainty relation,
\[
\Delta \varepsilon \; \tau = \hbar.
\]

STEP 2: If the measurement is performed over a cube of scale
$\sim T$, then there will be
\[
N \sim (T/\tau)^3
\]
cells.

STEP 3: The total uncertainty in energy is
\[
\Delta E = N \Delta \varepsilon = \hbar \frac{T^3}{\tau^4} .
\]

STEP 4: Schwarzschild radius associated with $\Delta E$ is
\[
R_{\rm Sch}=\frac{2G\, \Delta M}{c^2} = \frac{2G\, \Delta
E}{c^4} =
c \tau^2_{\rm p} \;  \frac{T^3}{\tau^4}.
\]

STEP 5: To avoid formation of the horizon, set
\[
R_{\rm Sch} < c \, T.
\]

STEP 6: This leads to the many-cell horizon limit,
\BEq
\tau > \sqrt{\tau_{\rm p} \, T}  \gg
\tau_{\rm p}.
\EEq

We estimate this for one of the most precise tests of
QED, the Lamb shift, which is on the order of 
\[
\Delta E_{\rm
Lamb} \sim 10^{-5} \; {\rm eV} \; \sim 10^{-24}\; {\rm J}.
\]
This corresponds to
\[
T_{\rm Lamb} \sim \hbar / \Delta E_{\rm Lamb} \sim 10^{-10} 
\;{\rm s}.
\]
The localization limit is then
\BEq
\tau_{\rm Lamb} > \sqrt{\tau_{\rm p} \, T_{\rm Lamb}}  \sim 
3 \times 10^{-27} \; {\rm s},
\EEq
which is 17 orders of magnitude greater than Plank's limit,
and about 100 times smaller than the chronon size associated
with the top quark. 

{\bf Conclusion: In the new theory the localization limit
must be present from the outset. This leads to the idea of
a chronon.}

\newpage

\section*{ \center APPENDIX E \\ \center THE SQUAD LEMMA}
\label{sec:SQUADLEMMA}

\addcontentsline{toc}{section}{APPENDIX E}

Here we reduce a squadron to a sequence of smaller squadrons.
This reduces a single Clifford-Wilczek quantification
to two quantifications in succession, one Clifford-Wilczek
and one Maxwell-Boltzmann. This means that one quantification
can replace the two needed for standard physics.

By a {\it squad} we mean a squadron whose top element
$\gamma^{\uar}$ has positive signature
(${\gamma^{\uar}}^2=+1$) and is
non-central.   A squad must have an even number of units.
The possible signatures depend on the number of units.
Any six independent units of
neutral total signature comprise a squad.
Any  eight independent units of any signature $\si$ comprise a
squad $\28_{\si}$. Eight is the least non-trivial number with this
property.
$N\28_{\si}$ is also a squad for any positive integer $N$.

If $\Bc,\Cc\subset \Ac$ are two subalgebras of the (real
associative unital) algebra $\Ac$, we define the {\it
product}  subalgebra $\Bc\, \Cc$ as the span of the set of
algebraic  products $\{bc
\,|\, b \in \Bc, c\in
\Cc\}$.

If $\Bc\cap\Cc=\21$ and
$\forall b\in \Bc \,\, \forall c\in \Cc
\,|\, bc=cb$,
then $\Bc \, \Cc=\Bc\ox\Cc$, the {\it tensor product} of the
two algebras.

\paragraph{Squad lemma} \cite{budinich, lounesto, porteous,
snygg} If $\3P$ is the mode space of a squad
then
\BEq
\Cliff(N {\3P}) \cong_{\dag}
\Ox_1^N \Cliff({\3P}).
\EEq

In other words, a Clifford product of $N$ squads
is algebraically $\dag$-isomorphic to a Maxwell-Boltzmann
sequence of those  squads.
In this way Clifford statistics naturally generates
Maxwell-Boltzmann
statistics for its squads.

Maxwell-Boltzmann statistics then reduces to
Fermi, Bose, and all the para- statistics.
This seems adequate for much of physics.

The construction of this isomorphism resembles the
well-known Jordan-Wigner construction of higher-dimensional
spin representations. We repeat it in the present context
for convenient  reference.

For definiteness we exhibit the construction for a neutral
hexad ${\bf 6_0}$. We label the
$6N$ generators $i_{\omega}(n)$ of
$\Cliff(N {\bf 6_0})$,
\BEq
\Cliff(N {\bf 6_0}) = \Cliff[i_5(N), \, \dots, \,
i_0(N), \, \dots, \, i_5(1), \, \dots, \,
i_0(1) ],
\EEq
     by two indices, an {\em internal} hexad index
$\omega =0, 1, 2, \dots, 5$ and an {\em external} hexad index
$n=1, 2, \dots , N$.

The generators $i_\omega (n)$ obey the usual Clifford
algebraic relations
\BEq
\{i_\omega (n), i_\rho (n') \} = +2 \delta (n, n')
G_{\omega \rho}(n),
\EEq
with
\BEq
(G_{\omega \rho}) = \left[ \matrix{ (g_{\mu\nu}) & 0 \cr 0
& (\delta_{\alpha\beta})} \right],
\quad
(g_{\mu\nu}) = \left[\matrix{ +1 & 0 & 0 & 0
\cr  0 & -1 & 0 & 0  \cr 0 & 0 & -1 & 0  \cr
0 & 0 & 0 & -1 }\right],
\quad
(\delta_{\alpha\beta}) = \left[\matrix{ 1 & 0  \cr 0 &1
}\right],
\EEq
and are symmetric with respect to the metric $G_{\omega
\rho}$:
\BEq
i_\omega (n)\adj = + i_\omega (n).
\EEq

Define the top element of each hexad,
\BEq
i^{\uar}(n) := i_5(n) \dots i_1(n) i_0(n).
\EEq
As we have already indicated,
\BEqA
\left( i^{\uar}(n) \right)^2 &=& + 1, \cr
\left( i^{\uar}(n) \right)\adj  &=& - i^{\uar}(n),\cr
[ i^{\uar}(n), i^{\uar}(n') ]  &=& 0,\cr
     [ i^{\uar}(n),
i_{\omega}(n') ]  &=& 0 \quad\mbox{ for }
     n \neq n' ,\cr
\{ i^{\uar}(n), i_{\omega}(n) \}  &=& 0.
\EEqA

We now define {\it local units} $\Gamma_\omega (n)$
as the Clifford
products
\BEq
\label{eq:LOCALGAMMA}
\Gamma_\omega (n) := (-1)^{n+1}\, i_\omega (n)
\prod_{m<n}^{\lar} i^{\uar} (m),
\EEq
ordered with $m$ increasing from left to right.
Then
\BEqA
[ \Gamma_{\omega}(n), \Gamma_{\rho}(n') ] &=& 0, \mbox{ for }
n \neq n' ;\cr
     \{ \Gamma_{\omega}(n), \Gamma_{\rho}(n) \}
&=&+2 G_{\omega \rho}(n),
\EEqA
and
\BEq
\Gamma_\omega (n) \adj = + \Gamma_\omega (n)
\EEq
is Hermitian with respect to $G_{\omega \rho}(n)$.

The local units generate the same
Clifford algebra
$\Cliff(N {\bf 6_0})$
     as the original pre-local units $i_{\omega }(n)$.
It follows then that
\BEq
\label{eq:HEXADLEMMA}
\Cliff(N {\bf 6_0}) \cong_{\dag} \underbrace{\Cliff({\bf
6_0}) \otimes
\dots
\otimes
\Cliff({\bf 6_0}) }_{N \; \rm times} \equiv
\bigotimes_{n=1}^N \Cliff({\bf 6_0}(n))
\EEq
as $\dag$-algebras. \hfill$\BOX$

This case of the squad lemma is the {\it hexad lemma}.
It
shows how a huge squadron of
pre-local anticommuting elementary processes can break up
into a Maxwell-Boltzmann sequence of commuting hexads of
local operations --- the seed of
classical space-time.
Similar results obtain for  any squad.

As a result, each term ${\bf 6_0}$ of $V=N{\bf 6_0}$ has a
Clifford algebra
$\Cliff(3,3)$ associated with it, whose spinors have eight
real components, forming an
$\28_0$.\footnote{Eight-component spinors have also been
used in physics by Penrose \cite{PENROSE}, Robson and
Staudte \cite{ROBSON}, and Lunsford \cite{LUNSFORD};
though not to unify spin with
space-time.} The
spinors of $V$ form the spinor space $\Sigma(N{\bf
6_0})=\Ox_1^N
\28_0=\28_0^N$.

\newpage

\section*{ \center  APPENDIX F \\ \center PROJECTIVE
REPRESENTATIONS OF THE PERMUTATION GROUP}

\label{sec:PROJREPSOFSYMMETRICGROUP}

\addcontentsline{toc}{section}{APPENDIX F}

Let us now turn to projective representations of the
symmetric (permutation) groups that have long been known to
mathematicians, but received little attention from
physicists. Such representations were overlooked in physics
much like projective representations of the rotation groups
were overlooked in the early days of quantum mechanics.

For convenience, following \cite{schur,
KARPILOVSKY, HOFFMAN} ({\it cf.} also \cite{WILCZEK,
wilczek}), we briefly
recapitulate the main results of Schur's theory.

One especially useful presentation of the
symmetric group $S_N$ on $N$ elements is given by
\BEqA
S_N &=& \bra \; t_1, \dots  , t_{N-1} \; : \; t_i^2 = 1,\;
(t_jt_{j+1})^3=1, \; t_kt_l=t_l t_k \; \ket ,\nonumber \\
&& 1\leq i
\leq N-1,\; 1\leq j \leq N-2, \; k\leq l-2 .
\EEqA
Here $t_i$ are transpositions,
\BEq
t_1=(12), t_2=(23), \dots , t_{N-1}=(N-1 N).
\EEq
Closely related to $S_N$ is the group $\tilde S_N$,
\BEqA
\label{eq:REPRESENTATIONGROUP}
\tilde S_N &=& \bra \; z, {t'}_1, \dots  , {t'}_{N-1}\; :\;
z^2=1,\; z{t'}_i = {t'}_i z,\;{t'_i}^2 = z,\;
({t'}_j{t'}_{j+1})^3=z, \; {t'}_k{t'}_l= z \, {t'}_l {t'}_k
\;
\ket ,\nonumber
\\ && 1\leq i
\leq N-1,\; 1\leq j \leq N-2, \; k\leq l-2 .
\EEqA

A celebrated theorem of Schur (Schur, 1911 \cite{schur})
states the following:

\noindent (i) The group $\tilde S_N$ has order $2(n!)$.

\noindent (ii) The subgroup $\{1,z\}$ is central, and is
contained in the commutator subgroup of $\tilde S_N$,
provided
$n\geq 4$.

\noindent (iii) $\tilde S_N / \{1,z\} \simeq S_N$.

\noindent (iv) If $N < 4$, then every projective
representation of
$S_N$ is projectively equivalent to a linear representation.

\noindent (v) If $N\geq 4$, then every projective
representation of
$S_N$ is projectively equivalent to a representation
$\rho$,
\BEqA
\rho(S_N) &=& \bra \; \rho(t_1), \dots ,
\rho(t_{N-1}): \rho(t_i)^2 = z, (\rho(t_j)\rho(t_{j+1}))^3=z,
\nonumber \\
&& \rho(t_k)\rho(t_l)= z \, \rho(t_l) \rho(t_k)
\; \ket ,
\nonumber \\ && 1\leq i
\leq N-1,\; 1\leq j \leq N-2, \; k\leq l-2 ,
\EEqA
where $z = \pm 1$. In the case $z = +1$, $\rho$
is a linear representation of $S_N$.

The group $\tilde S_N$ (\ref{eq:REPRESENTATIONGROUP}) is
called the {\it representation group} for $S_N$.

The most elegant way to construct a {\it projective}
representation
$\rho(S_N)$ of $S_N$ is by using the complex Clifford
algebra
${\Cliff}_{\Cbb}(V, g)\equiv \Cc_N$ associated with the
real vector space
$V=N\Rbb $,
\BEqA
\label{eq:CLIFFALGEBRA}
\{\gamma_i, \gamma_j\}=-2g(\gamma_i, \gamma_j).
\EEqA
Here $\{\gamma_i\}_{i=1}^N$ is an orthonormal basis
of $V$ with respect to the symmetric bilinear form
\BEq
g(\gamma_i, \gamma_j)=+\delta_{ij}.
\EEq
Clearly, any subspace $\bar V$ of $V=N\Rbb $ generates a
subalgebra
${\Cliff}_{\Cbb}(\bar V,\bar g)$, where $\bar g$ is the
restriction of
$g$ to $\bar V \times \bar V$. A particularly
interesting case is realized when $\bar V$ is
\BEq
\bar V:=\left\{ \sum_{k=1}^N \alpha^k \gamma_k \; : \;
\sum_{k=1}^N
\alpha_k=0\right\}
\EEq
of codimension one, with the
corresponding subalgebra denoted by
$\bar \Cc_{N-1}$ \cite{HOFFMAN}.

If we consider a special basis $\{t'_k\}_{k=1}^{N-1}\subset
\bar V$ (which is {\it not} orthonormal) defined by
\BEq
\label{eq:SWAPS}
t'_k:=\frac{1}{\sqrt{2}}(\gamma_k - \gamma_{k+1}), \quad
k=1, \dots, N-1,
\EEq
then the group generated by this basis is isomorphic to
$\tilde S_N$. This can be seen by mapping
$t_i$ to
${t'}_i$ and $z$ to -1, and by noticing that:

\noindent 1) For $k=1, \dots, N-1$:
\BEqA
{t'_k}^2 &=&\frac{1}{2}(\gamma_k - \gamma_{k+1})
(\gamma_k - \gamma_{k+1}) \nonumber \\
&=&  \frac{1}{2}(\gamma_k^2 + \gamma_{k+1}^2)
\nonumber \\
&=& - 1;
\EEqA

\noindent 2) For $N-2 \geq j$:
\BEqA
t'_jt'_{j+1}t'_j &=&\frac{1}{2\sqrt{2}}(\gamma_j -
\gamma_{j+1}) (\gamma_{j+1} - \gamma_{j+2})(\gamma_j -
\gamma_{j+1}) \nonumber \\
&=&  \frac{1}{\sqrt{2}}(\gamma_j - \gamma_{j+2}) ,
\EEqA
and
\BEqA
t'_{j+1}t'_jt'_{j+1}
&=&  \frac{1}{\sqrt{2}}(\gamma_{j+2} - \gamma_{j}) ,
\EEqA
so
\BEqA
(t'_jt'_{j+1})^3 &=&\frac{1}{2}(\gamma_j - \gamma_{j+2})^2
\nonumber \\
&=& -1 ;
\EEqA

\noindent 3) For $N-1\geq m>k+1$:
\BEqA
t'_kt'_m &=&+\frac{1}{2}(\gamma_k - \gamma_{k+1})
(\gamma_m - \gamma_{m+1}) \nonumber \\
&=& - \frac{1}{2}(\gamma_m - \gamma_{m+1}) (\gamma_k -
\gamma_{k+1})
\nonumber \\
&=& - t'_mt'_k .
\EEqA

One choice for the matrices is provided by the following
construction (Brauer and Weyl, 1935 \cite{brauer}):
\begin{eqnarray}
\label{eq:BRAUERWEYL}
\gamma_{2k-1} = \sigma_3 \otimes \cdots \otimes \sigma_3
\otimes
(i\sigma_1) \otimes {\bf 1} \otimes \cdots \otimes {\bf 1},
\cr \gamma_{2k} =
\sigma_3 \otimes \cdots \otimes \sigma_3 \otimes
(i\sigma_2) \otimes {\bf 1} \otimes \cdots \otimes {\bf 1}, \cr
k = 1, \, 2, \, 3,..., \, M,
\end{eqnarray}
for $N=2M$. Here $ \sigma_1$, $ \sigma_2$ occur in the
$k$-th position, the product involves $M$ factors, and
$\sigma_1$,
$\sigma_2$, $\sigma_3$ are the Pauli matrices.

If $N=2M+1$,
we first add one more matrix,
\BEq
\gamma_{2M+1} = i\,\sigma_3 \otimes \cdots \otimes \sigma_3
\quad (M \; {\rm factors}),
\EEq
and then define:
\begin{eqnarray}
\Gamma_{2k-1} &:=& \gamma_{2k-1}\oplus \gamma_{2k-1},
\nonumber \\
\Gamma_{2k} &:=&
\gamma_{2k}\oplus \gamma_{2k},
\nonumber \\
\Gamma_{2M+1} &:=&
\gamma_{2M+1}\oplus (-\gamma_{2M+1}).
\end{eqnarray}

The representation $\rho(S_N)$ so constructed is
reducible. An irreducible module of $\bar\Cc_{N-1}$
restricts that representation to the irreducible
representation of
$\tilde S_N$, since
$\{t'_k\}_{k=1}^{N-1}$ generates $\bar \Cc_{N-1}$ as an
algebra
\cite{HOFFMAN}.

\newpage

\section*{ \center APPENDIX G \\ \center APPLICATIONS TO THE
THEORY OF THE FQHE}
\label{sec:CLIFFORDOSCILLATOR}

\addcontentsline{toc}{section}{APPENDIX G}

In this Appendix I will describe my very first attempt at
understanding the Clifford statistics.
Using this statistics
I proposed a simple model for the grand canonical ensemble
of the carriers in the theory of fractional quantum Hall
effect. The model led to a
temperature limit associated with the
permutational degrees of freedom of such an ensemble.

As was pointed out before,
building on the work on nonabelions of Read and Moore
\cite{moore90, read92}, Nayak and Wilczek
\cite{NW, WILCZEK, wilczek}
proposed a spinorial statistics for the
fractional quantum Hall effect (FQHE) carriers. The
prototypical example was furnished by a so-called Pfaffian
mode (occuring at filling fraction $\nu = 1/2$), in which
$2n$ quasiholes form an
$2^{n-1}$-dimensional irreducible multiplet of the
corresponding braid group. The new
statistics was clearly non-abelian: it represented the
permutation group
$S_{N}$ on the $N$ individuals by a non-abelian group of
operators in the $N$-body Hilbert space, a projective
representation of $S_{N}$.

Since the subject is new, many unexpected effects in the
systems of particles obeying Clifford statistics may arise in
future experiments. One simple effect, which seems
especially relevant to the FQHE, might be observed in a grand
canonical ensemble of Clifford quasiparticles. Here I give
its direct derivation first.

Following Read and
Moore \cite{read92} we postulate that only two quasiparticles
at a time can be added to (or removed from) the FQHE
ensemble. Thus, we start with an $N=2n$-quasiparticle
effective Hamiltonian whose only relevant to our
problem energy level
$E_{2n}$ is
$2^{n-1}$-fold degenerate.
The degeneracy of the ground mode with no quasiparticles
present is taken to be
$g(E_0)=1$.

Assuming that
adding a {\it pair} of quasiparticles to the composite
increases the total energy by $\varepsilon$, and ignoring
all the external degrees of freedom, we can tabulate the
resulting many-body energy spectrum as follows:
\BEq
\begin{array}{|l|c c c c c c c c|}\hline
{\rm Number \, of \, Quasiparticles}, \, N= 2n   &  0 & 2 & 4
& 6 & 8 & 10 & 12 & \cdots \\ \hline
{\rm Degeneracy}, \, g(E_{2n}) = 2^{n-1} & 0 & 1 & 2 & 4 & 8
& 16 & 32 &
\cdots \\ \hline
       {\rm Composite \, Energy}, \,E_{2n}   &  0\varepsilon &
1\varepsilon & 2\varepsilon & 3\varepsilon & 4\varepsilon &
5\varepsilon & 6\varepsilon &  \cdots \\ \hline
\end{array}
       \EEq

Notice that the energy levels so defined furnish irreducible
multiplets for projective representations of permutation
groups in Schur's theory \cite{schur}, as was first pointed
out by Wilczek \cite{WILCZEK, wilczek}.

We now consider a grand canonical ensemble of Clifford
quasiparticles.

       The probability that the composite
contains $n$ pairs of quasiparticles, is
       \begin{eqnarray}
       P(n,T)=
       \frac{ g(E_{2n})e^{(n\mu-E_{2n})/k_BT} }{ 1 +
\sum_{n=1}^
       {\infty} g(E_{2n})e^{(n\mu-E_{2n})/k_BT} } \nonumber  \\
       \equiv \frac{ 2^{n-1} e^{-(n\mu-E_{2n}) /k_BT} }
       { 1 + \sum_{n=1}^{\infty}2^{n-1} e^{-n(\varepsilon -\mu)
/k_BT} },
\end{eqnarray}
where $\mu$ is the quasiparticle chemical potential.
       The denominator of this expression is the grand
partition function of the composite,
       \begin{eqnarray}
       Z(T)= 1 + \sum_{n=1}^{\infty}
g(E_{2n})e^{(n\mu -E_{2n})/k_BT} \nonumber \\
       \equiv 1 + \sum_{n=1}^{\infty}2^{n-1}
	 e^{-n(\varepsilon -\mu) /k_BT}
       \end{eqnarray}
       at temperature $T$.

       Now,
       \begin{eqnarray}
\sum_{n=1}^{\infty}2^{n-1}e^{-nx}=&e^{-x} \;[2^0 e^{- 0x}
       +2^1 e^{-1x} +2^2e^{-2x}+\cdots] \cr
       =& e^{-x} \; \sum_{n=0}^{\infty}e^{n(\ln2-x)}.
       \end{eqnarray}

       The partition function can therefore be written as
       \begin{eqnarray}
        Z(T) =1 + e^{- (\varepsilon -\mu) /k_BT}
       \sum_{n=0}^{\infty}e^{n(\ln2-(\varepsilon -\mu) /k_BT)}.
       \end{eqnarray}

    This leads to two interesting possibilities (assuming
$\varepsilon > \mu$):

       1) Regime $0<T<T_c$, where
       \begin{equation}\label{eq:TC}
       T_c= \frac{\varepsilon -\mu}{k_B\ln{2}}\/.
       \end{equation}

       Here the geometric series converges and
       \begin{eqnarray}
       Z(T) = \frac{ 1 - e^{- (\varepsilon -\mu) /k_BT} }
       { 1 - 2e^{- (\varepsilon -\mu)/k_BT} } =
\frac{ 2e^{-(\varepsilon -\mu)/2k_BT} }
{1-2e^{-(\varepsilon -\mu)/k_BT}} \; {\rm
sinh}\left( \frac{ \varepsilon -\mu}{ 2k_BT }\right).
       \end{eqnarray}

       The probability distribution is given by
       \begin{eqnarray}
       P(n, T) = \frac{  2^{n-1}e^{-n (\varepsilon -\mu) /k_BT}
       (1 - 2e^{- (\varepsilon -\mu)/k_BT})} {1 - e^{-
(\varepsilon -\mu) /k_BT} }.
       \end{eqnarray}

       2) Regime $T\geq T_c$.

       Under this condition the partition function diverges:
       \begin{equation}
       Z(T) = +\infty,
       \end{equation}
       and the probability distribution vanishes:
       \begin{equation}
       P(n,T) = 0 .
       \end{equation}

       This result indicates that the temperature $T_c$ of
(\ref{eq:TC}) is the upper bound of the intrinsic
temperatures that the
      quasiparticle ensemble can have. Raising the
temperature brings the system to higher energy levels which
are more and more degenerate, resulting in a heat capacity
that diverges at the temperature $T_c$\/.

To experimentally observe this
effect, a FQHE system should be subjected to a condition
where quasiparticles move freely between the specimen
and a reservoir, without exciting other degrees of freedom
of the system.

       A similar limiting temperature phenomenon seems to
occur in nature as
       the Hagedorn limit in particle physics \cite{hagedorn}.

Knowing the partition function allows us to find various
thermodynamic quantities of the quasiparticle system
for sub-critical temperatures $0<T<T_c$. We are particularly
interested in the average number of {\it pairs}
in the grand ensemble:
       \begin{equation}
       \label{eq:av.number}
       \bra n \ket_{\rm{Cliff}}= \lambda \frac{\partial \ln
Z}{\partial
\lambda},
       \end{equation}
       where $\lambda = e^{\mu /k_BT}$, or after
some algebra,
       \begin{eqnarray}
       \langle n(T) \rangle_{\rm{Cliff}} = \frac{e^{-
(\varepsilon -\mu) /k_BT}}
       {(1 - e^{-(\varepsilon -\mu)/k_BT})(1-2e^{-
(\varepsilon -\mu) /k_BT})} .
       \end{eqnarray}

       We can compare this with the familiar
Bose-Einstein,
       \begin{equation}
       \langle n(T) \rangle_{\rm{BE}} =
       \frac{1}{e^{(\varepsilon -\mu) /k_BT}-1} \equiv
       \frac{e^{- (\varepsilon -\mu) /k_BT}}{1 - e^{-
(\varepsilon -\mu) /k_BT}},
       \end{equation}
and Fermi-Dirac,
       \begin{equation}
       \langle n(T) \rangle_{\rm{FD}} =
       \frac{1}{e^{(\varepsilon -\mu) /k_BT}+1} \equiv
      \frac{e^{- (\varepsilon -\mu) /k_BT}}{1 + e^{-
(\varepsilon -\mu) /k_BT}},
       \end{equation}
distributions.
For the Clifford oscillator,
$\langle n(T) \rangle_{\rm{Cliff}} \rightarrow +\infty$ as
$T \rightarrow T_c -$, as had to be expected.

To relate Schur's theory of projective
representations of the permutation groups (descibed in
Appendix F) to physics we may try to
define a new, purely permutational variable of the Clifford
composite, whose spectrum would reproduce the degeneracy of
Read and Moore's theory.

A convenient way to define such a variable
is by the process of quantification.

Let us thus assume that if there is just one quasiparticle in
the system, then there is a limit on its localization, so
that the quasiparticle can occupy only a finite number of
sites in the medium, say $N=2n$. We further assume that the
Hilbert space of the quasiparticle is {\it real} and
$N=2n$-dimensional, and that a one-body
variable (which upon quantification corresponds to the
permutational variable of the ensemble) is an antisymmetric
generator of an orthogonal transformation of the form
\begin{eqnarray}
\label{eq:G}
G:= A
\left[\matrix{{\bf 0}_n & {\bf 1}_n \cr
-{\bf 1}_n& \hphantom{-}{\bf 0}_n}\right] ,
\end{eqnarray}
where $A$ is a constant coefficient. Note that in the
complex case this operator would be proportional to the
imaginary unit
$i$, and the corresponding unitary transformation would be a
simple multiplication by a phase factor with no
observable effect. Since the quantified operator algebra for
$N>1$ quasiparticles will be complex, the effect of just one
such ``real'' quasiparticle should be regarded as
negligible in the grand canonical ensemble.

In the non-interacting case the process of quantification
converts
$G$ into a many-body operator $\hat G$ by the rule
\begin{eqnarray}
\label{eq:hatG}
\hat{G}&:= &\sum_{l, j}^N
\hat e_l G^l{}_j \hat e^j,
\end{eqnarray}
where usually $\hat e_i$ and $\hat e^{j}$ are
creators and annihilators, but in more general situations
are the generators (that appear in the commutation or
anticommutation relations) of the many-body operator
algebra.

In Clifford statistics the generators of the
algebra are Clifford units $\gamma_i=2\,\hat e_i
=-\gamma_i\adj$, so
quantification of $G$ proceeds as follows:
\begin{eqnarray}
\label{eq:HATG}
\hat{G}
&= & -A \sum_{k=1}^{n}
(\hat e_{k+n}\hat e^k -
\hat e_k\hat e^{k+n}) \cr
&= & +A \sum_{k=1}^{n}
(\hat e_{k+n}\hat e_k -
\hat e_k\hat e_{k+n}) \cr
&= & 2 A \sum_{k=1}^{n}\hat e_{k+n}\hat e_k \cr
&\equiv &
\frac{1}{2} A \sum_{k=1}^{n} \gamma_{k+n}
\gamma_k.
\end{eqnarray}

Again, by Stone's theorem, the generator $\hat G$ acting on
the spinor space of the complex Clifford composite of $N=2n$
individuals can be factored into a Hermitian operator
$\tilde O$ and an imaginary unit
$i$ that commutes strongly with $\tilde O$:
\BEq
\hat G = i \tilde O.
\EEq
We suppose that $\tilde O$ corresponds to the permutational
many-body variable mentioned above, and seek its spectrum.

We note that $\hat G$ is a sum of $n$
commuting anti-Hermitian algebraically independent operators
$\gamma_{k+n} \gamma_k, \; k=1,2,...,n$,
$(\gamma_{k+n} \gamma_k)^{ \dagger }=-\gamma_{k+n}
\gamma_k$,
$(\gamma_{k+n} \gamma_k)^2=-1$.
If we now use $2^n \times 2^n$
complex matrix representation of Brauer and Weyl
(\ref{eq:BRAUERWEYL}) for the
$\gamma$-matrices, we can simultaneously
diagonalize the $2^n\times2^n$ matrices representing the
commuting operators $\gamma_{k+n} \gamma_k$, and use their
eigenvalues, $\pm i$, to find the spectrum of
$\hat G$, and consequently of $\tilde O$. The final result
is obvious: there are $2^{2n}$ eigenkets of $\tilde
O$, corresponding to the dimensionality of the spinor
space of
${\Cliff}_{\Cbb}(2n)$. In the irreducible
representation of
$S_N$ this number reduces to $2^{n-1}$, as
required by Read and Moore's theory.

Note that in this approach the possible number of the
quasiholes in the ensemble is fixed by the number of
the available sites, $N=2n$. A change in that
number must be accompanied by a change in the
dimensionality of the one-quasiparticle Hilbert space.
It is natural to assume that variations in the physical
volume of the entire system would privide such a mechanism.

\end{document}